\documentclass[aps,prb,reprint,amsmath,amssymb,showpacs,floatfix,superscriptaddress]{revtex4-2}

\usepackage{graphics}                  
\graphicspath{ {./figures/} }
\usepackage{color}
\usepackage{changes}

\definecolor{red}{rgb}{1.00,0.00,0.00}
\usepackage{todonotes}

\usepackage{xspace}
\renewcommand{\deg}{$^{\circ}$\xspace}
\newcommand{\cso}{Cu$_2$OSeO$_3$\xspace}

\begin{document}

\title{
Time-resolved measurement of spin excitations in Cu$_2$OSeO$_3$
}

\author{D. M. Burn}
\affiliation{Diamond Light Source, Harwell Science and Innovation Campus, Didcot, OX11~0DE, United Kingdom}

\author{S. L. Zhang}
\affiliation{School of Physical Science and Technology, ShanghaiTech University, Shanghai 200031, China}
\affiliation{ShanghaiTech Laboratory for Topological Physics, ShanghaiTech University, Shanghai 200031, China} 

\author{G. \surname{van der Laan}}
\affiliation{Diamond Light Source, Harwell Science and Innovation Campus, Didcot, OX11~0DE, United Kingdom}
	
\author{T. Hesjedal}
\affiliation{Clarendon Laboratory, Department of Physics, University of Oxford, Parks Road, Oxford, OX1~3PU, United Kingdom}

\begin{abstract}
Magnetic diffraction in combination with x-ray detected ferromagnetic resonance (DFMR) is a powerful technique for performing time-resolved measurements on individual spin textures.
Here, we study the ferromagnetic resonance (FMR) modes of both the conical and field-polarized phases in the chiral magnet \cso.
Following the identification of the FMR modes at different temperatures using broadband vector network analyzer FMR, we use DFMR on the crystalline (001) Bragg peak to reveal the time-dependent spin configurations of the selected FMR modes.
By being able to measure both the amplitude and phase response of the spin system across the resonance,  a continuous phase advance (of $180^\circ$) in the conical mode, and a phase lag (of $180^\circ$) in the field-polarized mode is found.
By performing dynamic measurements in the conical phase as a function of the linear polarization angle of the x-rays, i.e., successively probing the dynamics of the moments, we find an inversion of the dynamics along the conical axis upon inverting the applied field direction.
By allowing for time-resolved measurements of the phase and amplitude of individual magnetic phases, DFMR opens up new opportunities for obtaining a deeper understanding of the complex dynamics of chiral magnets.
\end{abstract}

\date{\today}
\maketitle


\section{Introduction}

Topological spin textures in chiral magnets such as Cu$_2$OSeO$_3$ are of great interest for potential advances in spintronics \cite{Onose_PRL_2012, NatNano_8_899}.
Key to the success of future technologies are their dynamic properties, enabling fast and energy-efficient transitions of the magnetic states.
One of the most commonly used techniques for the study of the magnetization dynamics is ferromagnetic resonance (FMR) \cite{Cochran1986,Metaxas2010,Magaraggia2011,Kaiser2011}.
In broadband coplanar waveguide (CPW) FMR, the resonances of a sample are probed by placing it on the CPW and subjecting it to a variable magnetic field.
By sweeping an RF signal fed to the CPW in the range from typically 0 to 20\,GHz, and by measuring the transmitted RF power, the resonances reveal themselves as absorption maxima in frequency vs field maps.

Previous vector network analyzer (VNA) FMR studies have probed the magnetization dynamics of the resonance modes in Cu$_2$OSeO$_3$, i.e., the helical, conical, and skyrmion phase, as well as the field-polarized phase, revealing their evolution in field-frequency space \cite{Onose_PRL_2012, Schwarze_NMat_2015, Stasinopoulos_SciRep_2017, Weiler_PRL_2017, Garst2017}. 
Supported by micromagnetic simulations, three internal breathing and rotational modes have been identified in the skyrmion phase \cite{Onose_PRL_2012}, two each in the helical and conical states, and one in the field-polarized state \cite{Schwarze_NMat_2015}.

Within the conical phase, the two resonance modes that have been resolved can be attributed to vibrational modes in the conical structure and are labeled $+Q$ and $-Q$ \cite{Onose_PRL_2012}.
While their magnetization oscillates uniformly, the two modes have opposite helicity and oscillate (counter)clockwise in the ($+Q$) $-Q$ mode \cite{Garst2017}.
For their excitation, the AC field has to be perpendicular to the bias field direction, and hence the conical propagation axis \cite{Onose_PRL_2012}.
As well as the geometry of the field direction, the excitation also depends on the sample geometry due to the anisotropy \cite{Schwarze_NMat_2015, Stasinopoulos_SciRep_2017}. 
The Gilbert damping parameter is small and was found to be $<$0.003 in \cso\ at 5\,K (from the FMR linewidth) \cite{Weiler_PRL_2017}.
Note that the counterclockwise oscillating $+Q$ mode seamlessly connects to the also counterclockwise oscillating Kittel mode present in the field-polarized phase above a critical field, while the spectral weight of the $-Q$ mode is vanishing.
The counterclockwise rotation sense in the field-polarized state is a result of the fact that single electron moments, representative of a uniform magnetic state, simply precess counterclockwise around a magnetic field vector.

Resonant elastic x-ray scattering (REXS) provides a unique, element-specific view of ordered magnetic structures, and has been extensively used for the study of \cso\ \cite{PRB_93_214420, NanoLett_16_3285,CSO-APL_2016,NatComms_8_14619, Ukleev_2022} and other chiral magnetic systems \cite{Duerr1999, Zhang_PRB_2017, PhysRevLett.120.037202, Legrand2018, Li2019, Kim2022}.
The use of soft x-rays (via x-ray magnetic circular dichroism) to measure magnetization dynamics through x-ray detected ferromagnetic resonance (XFMR) is a valuable technique \cite{Arena2009,Marcham2011,bailey_XFMR,VanDerLaan_JESRP_2017}, giving access to the dynamics of individual layers via the element-selectivity of x-ray spectroscopy in few-layer stacks such as spin valves \cite{Baker2015} and magnetic tunnel junctions \cite{Baker2016}.
The combination of the two, REXS and XFMR, allows for dynamic studies of diffraction-selected specific magnetic modes, and is particularly useful for systems in which the eigenmodes are complex and their interpretation crucially relying on micromagnetic simulations.
The capabilities of this diffractive FMR (DFMR) technique was demonstrated on the rich magnetic structure of a Y-type hexaferrite \cite{Burn2020-DFMR}, and it was further expanded into the probing of superlattices via x-ray reflectometry \cite{PhysRevLett.125.137201}.
A similar technique, REXS-FMR, was successfully applied for the study of, most notably, the three skyrmion eigenmodes in \cso\ via the crystalline (001) Bragg peak and its magnetic satellites \cite{Back-DFMR}.
While REXS-FMR is a homodyne detection technique which relies on the difference of probed magnetic peak intensities between RF on and off,
heterodyne detection DFMR is a truly time-resolving technique, giving access the phase of the dynamic response \cite{Burn2020-DFMR}.
Also note that both REXS-FMR and DFMR can be carried out on the magnetic satellites as well, however, due to the complexity of the analysis, the present study will focus on dynamic Bragg peak studies.

Here, we report a time-resolved dynamic study of the conical and field-polarized phase in \cso\ using the diffractive ferromagnetic resonance technique, employing both circularly and linearly polarized x-rays.
First, the dynamic phase diagram is mapped out using VNA-FMR with the field applied in-plane, identifying the conical and field-polarized modes and characterizing their temperature and field dependence.
The characteristics of the resonance modes in the conical and field-polarized phase, as well as the dependence of the critical field on temperature, is well-captured using the established models \cite{Garst2017}.
Using DFMR, we investigate the dynamic behavior of the conical $+Q$ and the field-polarized mode, as well as the transition between them, in greater detail.
As the field-polarized phase does not result in magnetic scattering peaks, we restrict our study to the analysis of the (001) Bragg peak, whose magnetic intensity is susceptible to both phases.
Note that, in principle, any reflection peak could be probed which has a magnetic contribution.
We find a 180\deg phase shift across the resonances of the conical and field-polarized modes, which is negative and positive, respectively.
%
Further, while the conical resonance peak position is temperature-dependent, the field-polarized resonance is temperature-independent.
Finally, by employing linear polarized x-rays of variable polarization angle, we are able to access the dynamics of the projected moments along the conical texture at resonance.
We find that upon reversing the applied field, the dynamic response is inverted.


\section{Experimental}

\begin{figure}[tbh]
	\centering
	\includegraphics[width=8.6cm]{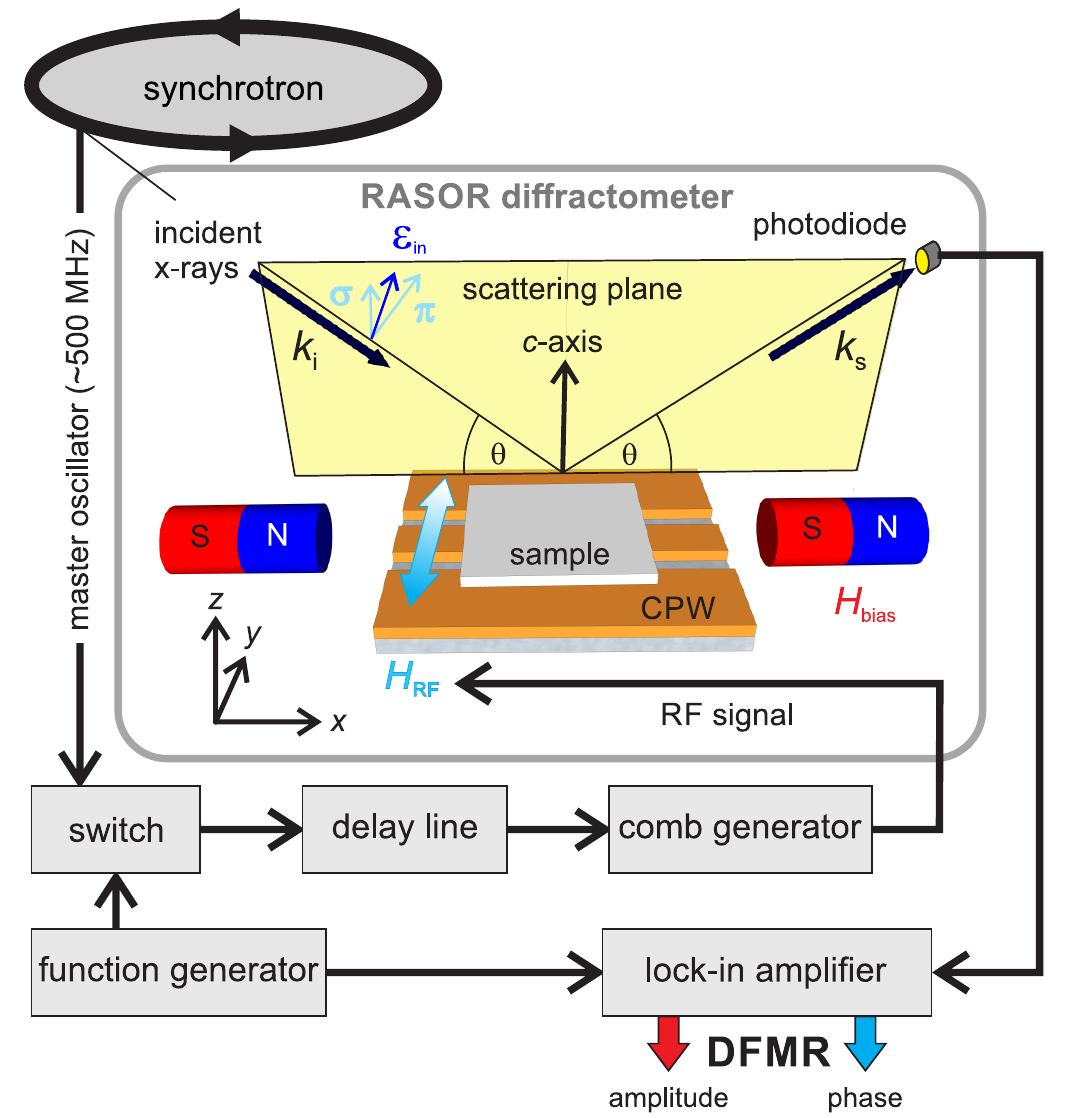}
	\caption{
		Schematic of the ferromagnetic resonance setup, allowing for both VNA-FMR and DFMR measurements in the RASOR diffractometer at the Diamond Light Source.
		The \cso\ sample is placed on the CPW, which is mounted on the cold finger inside the diffractometer.
		The incident x-rays, which are tuned to the Cu $L_3$ edge, are scattered off the sample and detected via a photodiode in a $\theta$-$2\theta$ geometry.
		Circularly as well as linearly polarized x-rays were used, whereby the linear polarization $\epsilon_i$ was continuously variable, covering $\epsilon$ being in the scattering plane ($\pi$) and perpendicular to it ($\sigma$). 
		A linear polarization angle of $0^\circ$ corresponds to linear horizontal ($\sigma$) polarization.
		The variable magnetic field is applied in the scattering plane via pairs of permanent magnets whose distance can be controlled.
		An RF signal is fed to the CPW to drive the ferromagnetic resonance in the magnetic sample, with its field effectively being perpendicular to the scattering plane at the probed sample position.
		As the synchrotron conveniently produces x-ray pulses at a frequency of $\sim$500\,MHz, a comb generator is used to produce higher harmonics, which are selected and fed to the CPW. 
		To probe the time-dependence of the scattered x-ray intensity, a tunable delay line is used, which shifts the phase between the pump (the RF signal) and the probe (the pulsed x-rays).
	}
	\label{setup}
\end{figure}

The dynamic behavior of the conical and field-polarized phase in \cso\ was measured using broadband VNA-FMR \cite{BroadbandFMR} and synchrotron-based DFMR \cite{Burn2020-DFMR}.
Both techniques were performed in-situ in the RASOR diffractometer on beamline I10 at the Diamond Light Source (Oxfordshire, UK).
A high-quality, (001)-cut \cso\ sample, measuring 2~$\times$~2~$\times$~0.1~mm$^3$, was polished and placed on a coplanar waveguide (CPW).
The CPW was attached to a cold-finger cryostat at the center of the diffractometer, and cooled to temperatures in the range from 25 to 60~K in ultrahigh vacuum. 
A magnetic bias field was applied in the plane of the sample by way of two permanent magnets, whereby the field strength can be controlled by varying the separation between the magnets. 
This bias field was oriented in the scattering plane, and perpendicular to the radio frequency (RF) field produced by the CPW. 
The schematic of the experimental setup is illustrated in Fig.\ \ref{setup}. 

First, the static magnetic scattering from the sample was characterized. 
\cso [cubic space group $P$2$_1$3 (198)] has a lattice constant of 8.925\,\AA.
At the Cu $L_3$ edge (931.25\,eV), this results in the (001) structural Bragg peak at $2\theta = 96.5^\circ$ \cite{CSO-APL_2016}.
Note that this (001) Bragg peak is `forbidden' and only observed in resonant x-ray scattering \cite{PRB_93_214420, NanoLett_16_3285, Templeton1994, Dmitrienko2012}.
In resonant scattering, the magnetic state of \cso\ can be accessed via the structural Bragg peak \cite{Back-DFMR}, i.e., effectively probes the $k=0$ dynamics, or, alternatively the magnetic satellites ($k \neq 0$ modes, e.g., the skyrmion breathing mode) \cite{PRB_93_214420, NanoLett_16_3285}.

To cover an extended field range, the measurements were performed in two parts, whereby the smaller permanent magnets were replaced by larger ones for the second part of the measurements.
As their field ranges were not perfectly overlapping, fields around 30~mT could not be covered (white areas in the resonance plots in Fig.\ \ref{vnafmr}).

As a first step, the dynamic phase space of the system was mapped out using VNA-FMR. 
The RF power transmitted through the CPW, $S_{21}$, is measured, whereby its loss ($\Delta S_{21}$) as a function of frequency represents the FMR spectrum. 
In VNA-FMR measurements, this absorption is typically mapped out as a function of RF frequency and magnetic bias field. 
However, in this investigation, the absorbed RF power also depends on the magnetic phase of the sample, and thus the excitable magnetic spin configurations, resulting in a temperature and extra field dependence of the spectra.
Therefore, in this investigation, we map out the RF absorption of the sample as a function of all three quantities, i.e., frequency, field, and temperature.
In practice, this is best achieved by performing frequency sweeps while slowly decreasing the temperature.
The difference in sweep rates of both quantities means that the temperature deviation over the time taken to measure a frequency sweep is negligible. 
These measurements are then repeated at various magnetic fields. 
This effectively measures the dynamic properties of the system as prepared using a field-cooling protocol.

As the temperature was varied continuously while taking frequency sweeps, the temperature steps need to be interpolated to enable further processing and the analysis of two-dimensional slices of frequency-field at fixed temperature or field-temperature at fixed frequency. 
In VNA-FMR measurements, the background resulting from the electrical characteristics of the measurement setup is subtracted.
This background is typically obtained from a high-field measurement where the moments are saturated and any dynamic features are driven to sufficiently high frequencies to be out of the measurement range. 
Due to the limitations in the field range in the RASOR diffractometer, we adopt an alternative approach, by subtracting a background measured in the field polarized state at 60\,K, i.e., at a temperature above $T_\mathrm{C}$.

After the detection of the dynamic modes using VNA-FMR, the modes were further investigated using polarized soft x-rays. 
Both circularly and linearly polarized x-rays were used, with an energy tuned to the Cu $L_3$ edge at 931.25\,eV.
The beam spot size at the sample position is typically 100 $\times$ 100~$\mu$m$^2$, and the scattered beam was measured using a photodiode in a $\theta$-$2\theta$ configuration with slits providing an angular resolution of $\sim$3\,mrad. 

Initially measurements of the scattered intensity as a function of reciprocal lattice space coordinates were used to map out the static magnetic phase diagram of the sample and to optimize the alignment of the diffractometer. The detector was then positioned at the (001) Bragg peak of \cso.

The polarized soft x-rays provide a probe of the magnetization state in the sample due to circular and linear magnetic dichroism in resonant elastic x-ray scattering \cite{VANDERLAAN2008570}.
As noted above, in resonant scattering on \cso, both the (001) Bragg peak \cite{Back-DFMR} as well as its surrounding magnetic satellites \cite{PRB_93_214420, NanoLett_16_3285} can be used to probe the magnetic states of \cso.
Here, due to the complexity of depth-dependent information a resonant soft x-ray study of the magnetic satellites provides, we focus on the analysis of the (001) Bragg peak data.

For the dynamic measurements, the scattered intensity at the (001) Bragg peak was measured stroboscopically, taking advantage of the pulsed nature of the synchrotron radiation at a frequency of $\sim$500~MHz.  
Here, the RF magnetic field generated from the CPW beneath the sample was phase-locked to an integer multiple of the synchrotron master clock.  
Due to the reduced thickness of the sample ($\sim$100\,$\mu$m), the RF power delivered from the CPW was sufficient to excite the magnetization precession dynamics on the \cso\ top surface, probed by the x-rays.
The intensity of the scattered x-rays provides a measure of the magnetic state of the system for a particular point in time during the precession of the magnetization.
Taking such snapshots for increasing delay time between the RF pump and the x-ray probe are then giving the time-dependent magnetization behavior. 
The signal-to-noise ratio in the dynamic component of the scattered intensity is further enhanced through lock-in detection, using a 180\deg modulation of the phase of the RF pump at a frequency of 700\,Hz.
For our measurements, the RF frequency was set to 4\,GHz where a clear FMR mode crossing is found for both the conical and field polarized phases. 
Both of these mode crossings are within the field range accessible in the diffractometer. 
Measurements of the dynamic signal as a function of pump-probe delay and magnetic field were performed at various temperatures with both left and right circularly polarized x-rays, and also with linearly polarized x-rays as a function of the incident polarization angle.


\section{Results}

\begin{figure*}[tbh]
	\centering
	\includegraphics[width=178mm]{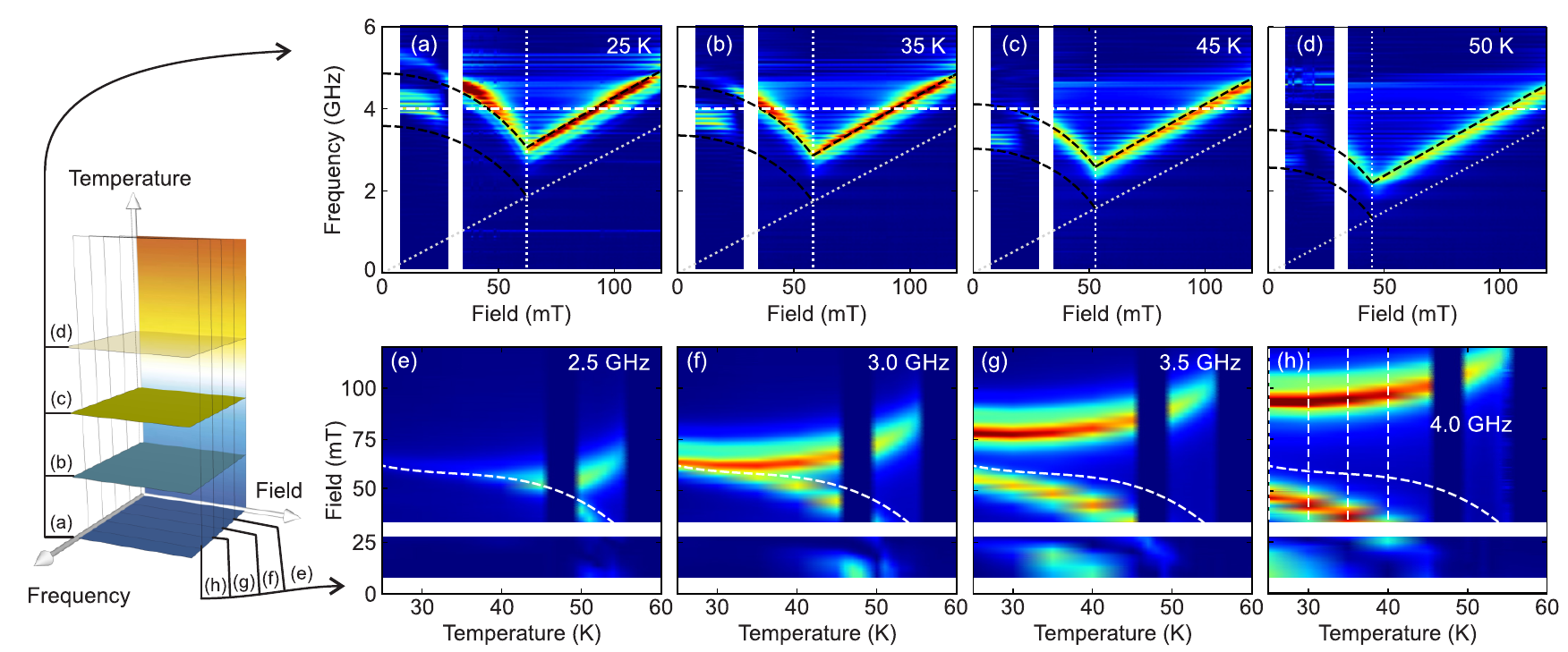}
	\caption{Vector network analyzer ferromagnetic resonance measurements showing  plots of the FMR spectra ($\Delta S_{21}$, linear scale in arbitrary units) for different measurement conditions.
	The data was obtained from a series of field-cooling temperature sweeps at different fields, with the field applied perpendicular to the RF excitation direction. 
	Slices of the data show the absorption as a function of (a-d) frequency and field, and (e-h) field and temperature.
	The strong absorption features represent magnetic resonance modes in the sample. The vertical dashed line (a-d) and curved lines (e-h) represent the magnetic phase transition field, $H_{c2}$, between the conical and field polarized phase. 
	Model plots of the resonance curves for both phases (black dashed lines) are overlaid on top of the data in panels (a-d), and the dotted white diagonal line represents the behavior of the Kittel mode in a perfect sphere.
	The horizontal dashed line at 4\,GHz (a-d), and the vertical dashed lines at 25, 30, 35, and 40\,K in (h) indicate the parameters for which the DFMR measurements were carried out.
	}
	\label{vnafmr}
\end{figure*}

The ferromagnetic resonance modes of the \cso\ sample were first studied in the RASOR diffractometer using VNA-FMR.
The frequency spectra were measured during a series of field-cooling sweeps, i.e., as a function of temperature. 
Slices of this data as a function of frequency and field at different temperatures, and also as a function of field and temperature at different frequencies are shown in Fig.\ \ref{vnafmr}(a-d) and Fig.\ \ref{vnafmr}(e-h), respectively. 
The color scale represents the magnitude of transmitted RF power, $S_{21}$, revealing strong absorption features corresponding to magnetic resonance modes in the sample. 

At the critical field $H_{c2}$, an abrupt change in the magnetic resonance behavior is observed, which is indicated in Fig.\ \ref{vnafmr}(a-d) by the vertical, and in Fig.\ \ref{vnafmr}(e-h) by the curved dashed lines.
This phase boundary separates the conical phase at lower fields from the field-polarized phase which exists above the transition.
Both of these phases show different magnetic resonance behavior and there is a continuous transition between them crossing the phase boundary.

The resonance behavior of the field-polarized mode above $H_{c2}$ is adequately described by the Kittel equation, i.e., the description for a uniform ferromagnet (see Eq.\,(14) in Ref.\ \cite{Garst2017}).
Note that while the sample in the present case is a flat plate with its surface normal along the $z$-direction (i.e. with demagnetization factors of $N_x \approx N_y \approx 0$ and $N_z \approx 1$), the static magnetic field is applied along the $x$-direction, while the RF field, $H_\mathrm{RF}$, is applied along the $y$-direction (see Fig.\ \ref{setup}).
The measured resonance curves [Fig.\ \ref{vnafmr}(a-d)] lie above the straight (white dotted) frequency-field line representing the behavior of a perfect sphere, i.e., when demagnetization is irrelevant (demagnetization factors $N_x=N_y=N_z=1/3$) and $\hbar \omega = g \mu_\mathrm{B} \mu_0 H$, which is the resonance frequency in the paramagnetic limit.
Note that the field-polarized phase branch can either lie above (as here) or below the paramagnetic resonance (dotted white lines), depending on whether the shape anisotropy effectively enhanced or reduces the applied static field \cite{Gurevich}.

From fitting the Kittel equation to the data [dashed black lines above $H_{c2}$ in Fig.\ \ref{vnafmr}(a-d)], the following parameters were obtained: 
for the demagnetization factors, $N_x=N_y=0.0012$ and $N_z = 0.9976$, and for the internal susceptibility, $\chi_\mathrm{con}^\mathrm{int} = 1.7$. This value for $\chi_\mathrm{con}^\mathrm{int} = \mu_0 M_s^2 / (2 A Q^2)$, where $M_s$ is the saturation magnetization, $Q$ is the pitch vector, and $A$ is the stiffness, is consistent with the value of 1.76 reported in Ref.\ \cite{Garst2017}.

The conical phase, which exists below $H_{c2}$, has a more complex dynamic behavior with two modes, $+Q$ and $-Q$, corresponding to the two vibrational modes within the chiral spin texture \cite{Onose_PRL_2012,Schwarze_NMat_2015}. 
The conical modulation period in \cso\ is between 50 and $\sim$60\,nm \cite{Seki_PRB_2012,Seki_Science_2012}.
Both modes show a decrease in frequency with increasing field, and are separated by a difference in frequency. 
At low field, the lower frequency $-Q$ mode contributes to a large RF absorption. With increasing field, the $-Q$ mode decreases in intensity and the $+Q$ mode becomes the dominant mode, in agreement with the theoretically calculated behavior \cite{Garst2017}. 
At $H_{c2}$, the $+Q$ mode continuously transforms into the field-polarized Kittel mode \cite{Stasinopoulos_SciRep_2017, Schwarze_NMat_2015, Garst2017}.
These dashed black lines in Fig.\ \ref{vnafmr}(a-d) represent fits to the data using the expression given by Garst \emph{et al.}\ in Eq.\,(30) of Ref.\,\cite{Garst2017}.

Further, the resonance modes have a subtle temperature dependence when traversing the field-temperature phase diagram [Figure \ref{vnafmr}(e-h)].
In the field-polarized phase, the resonance field experiences a slight increase up to $T_\mathrm{C}$ = 57\,K, above which the intensity of the mode vanishes. In the conical phase, the resonant field of the $+Q$ and $-Q$ modes both decrease with increasing temperature at fixed frequencies.


The VNA-FMR measurements shown in Fig.\ \ref{vnafmr} reveal the behavior of the resonance modes within the frequency-temperature-field phase space in \cso.
However, while the presence of power-absorbing resonances is clear from these measurements, nothing can be learned about their nature.
To gain further insight into their time-dependent spin structure, the dynamic contribution to the scattering at the (001) Bragg peak was measured while exciting specific ferromagnetic resonance modes.

\begin{figure}[tbh]
	\centering
	\includegraphics[width=86mm]{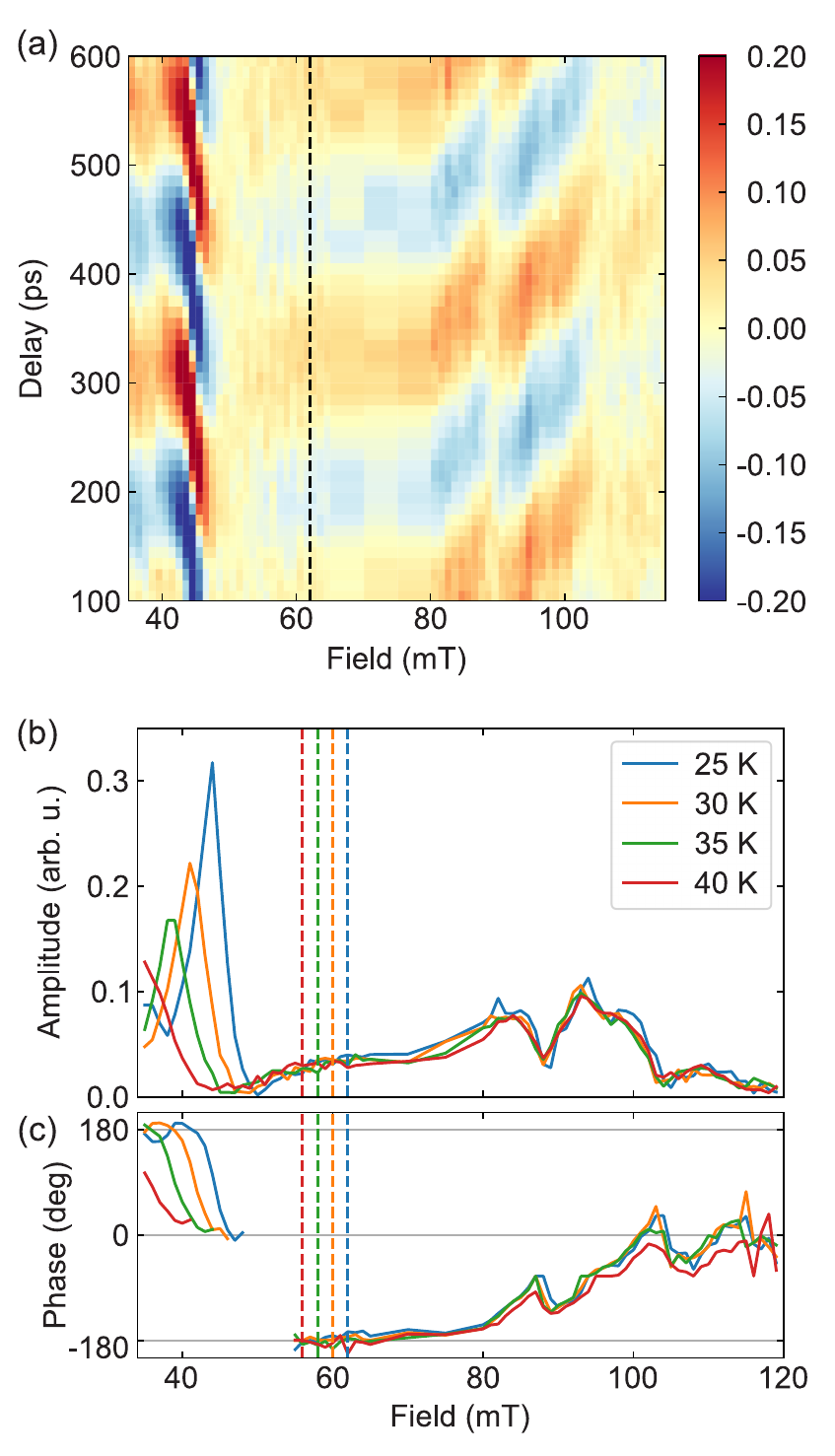}
	\caption{(a) Colormap showing the time dependent contribution to the (001) Bragg intensity, plotted as a function of the magnetic field and the delay between the RF pump and x-ray probe. 
	The measurements were done with an excitation frequency of 4\,GHz, at a temperature of 25\,K with the field applied in-plane and with right circularly polarized x-rays at the Cu $L_3$ edge.
	The color scale represents the value of the time-dependent contribution to the  (001) Bragg peak signal (in arbitrary units).
	The amplitude and phase data extracted from sinusoidal fits to the delay scans as a function of field are shown in panels (b) and (c), respectively.
	The vertical lines indicate $H_{c2}$ for the respective temperature, separating the conical and field polarized magnetic phases.
}
	\label{field_delay_amp_phase}
\end{figure}

During resonance, the precession of the magnetic moments about the equilibrium spin configuration results in a time-dependent variation of the projection of the moments along the direction of the x-ray polarization vector. 
This time-dependent projection in turn results in a time-dependent contribution to the scattered intensity at the (001) Bragg peak.
Figure \ref{field_delay_amp_phase}(a) shows a colormap plot of sequences of such time-dependent scattering intensities for an excitation frequency of 4\,GHz. 
The dynamic contributions to the scattering were recorded as a function of the time delay between the RF pump and x-ray probe pulse, and these so-called delay scans were further obtained as a function of external bias field.
The yellow regions indicate that the dynamic scattering is equal to the scattered intensity from the static state, while red (blue) regions indicate an increase (decrease) in the scattered intensity, respectively. 
The delay scans show sinusoidal oscillations with a period of 250\,ps, and demonstrate that the dynamics in the spin configuration is coupled to the 4\,GHz excitation frequency.

From the contrast in Fig.\ \ref{field_delay_amp_phase}(a) it can be seen that the amplitude and the phase of the oscillations of the (001) Bragg peak intensity change as a function of the applied magnetic field. A strong enhancement of the contrast is observed at 44\,mT, and a comparably weaker enhancement between 80 and 100\,mT, corresponding to the two resonance modes observed in the 4\,GHz VNA-FMR measurements at 25\,K, shown in Fig.\ \ref{vnafmr}(a). 
From the shifts of the patterns across Fig.\ \ref{field_delay_amp_phase}(a), especially around the fields at which the contrast is strong, it can also be concluded that the phase of the sinusoidal dynamic signal strongly depends on the applied magnetic field.


The dynamic contribution to the (001) Bragg intensity shown in Fig.\ \ref{field_delay_amp_phase}(a) can be quantified via sinusoidal fits to the delay scan data at each field point.
Figures \ref{field_delay_amp_phase}(b) and \ref{field_delay_amp_phase}(c) show the corresponding amplitude and phase, respectively, as a function of the magnetic field at 25, 30, 35, and 40\,K.
%
For all four temperatures, a peak is visible in the amplitude at $\sim$40\,mT, corresponding to the resonance mode in the conical phase.
These peaks are accompanied by a $-180$\deg phase change across the resonance.
Both the magnitude and the position of the amplitude peaks are a function of temperature, with the amplitude and resonant field being largest at low temperature, and decreasing with increasing temperature.
Both of these features are consistent with the VNA-FMR measurement data presented in Fig.\ \ref{vnafmr}(a-d).

The amplitude [Fig.\ \ref{field_delay_amp_phase}(b)] also shows features in the field range from 80-100\,mT, consisting of a more complex peak structure, which is also accompanied by a (positive) phase change of 180\deg [Fig.\ \ref{field_delay_amp_phase}(c)].
These features correspond to the resonance of the field-polarized phase and they show, in contrast to the conical mode, no apparent dependence on temperature in either peak amplitude or resonance field.
This behavior is consistent with the flatness of the resonance over this temperature range, which can be seen in Fig.\ \ref{vnafmr}(h).

Note that the +180\deg phase shift over the field-polarized resonance with field is opposite to the $-180$\deg phase shift over the conical resonance.
This leading or lagging behavior of the phase, 
is due to the different ways in which the resonance is crossed, i.e., with a negative frequency-field slope in case of the conical and a positive slope in case of the field-polarized resonance [see Fig.\ \ref{vnafmr}(a) for comparison].
Note that the results shown in Fig.\ \ref{field_delay_amp_phase} are measured with right-circularly polarized light with the field applied in-plane in the same orientation. When the field direction is reversed, the dynamic signal still shows a peak in amplitude for the conical resonance and the associated 180\deg phase change still occurs in the same direction.

The field-polarized resonance is rather broad and shows a detailed fine structure in both amplitude and phase (Fig.\ \ref{field_delay_amp_phase}).
This behavior is due to the existence of multiple magnetostatic modes in the sample, which are enhanced due to the thin thickness of the sample \cite{Storey_1977}. 
These multiple overlapping resonance modes combine, resulting in the complex feature in which the higher frequency-field component is the strongest contribution.
In the conical phase, these magnetostatic modes are also present, but occur over a narrower frequency range and are not resolved.
Note that these modes are also present in VNA-FMR measurements in both the field-polarized and conical phase, however, hard to discern in the usual representation of VNA-FMR absorption maps (they become visible upon differentiation of the signal). 
%
Further, the low-field tail of the field-polarized resonance mode appears to `leak' into a field range below the $H_{c2}$ transition, i.e., into the conical phase range.
The temperature-dependent $H_{c2}$ values are indicated by the vertical dashed lines in Fig.\ \ref{field_delay_amp_phase}(b,c).
This behavior points towards the existence of metastable states, resulting from the field-cooling protocol employed in our measurements.


\begin{figure}[tbh]
	\centering
	\includegraphics[width=86mm]{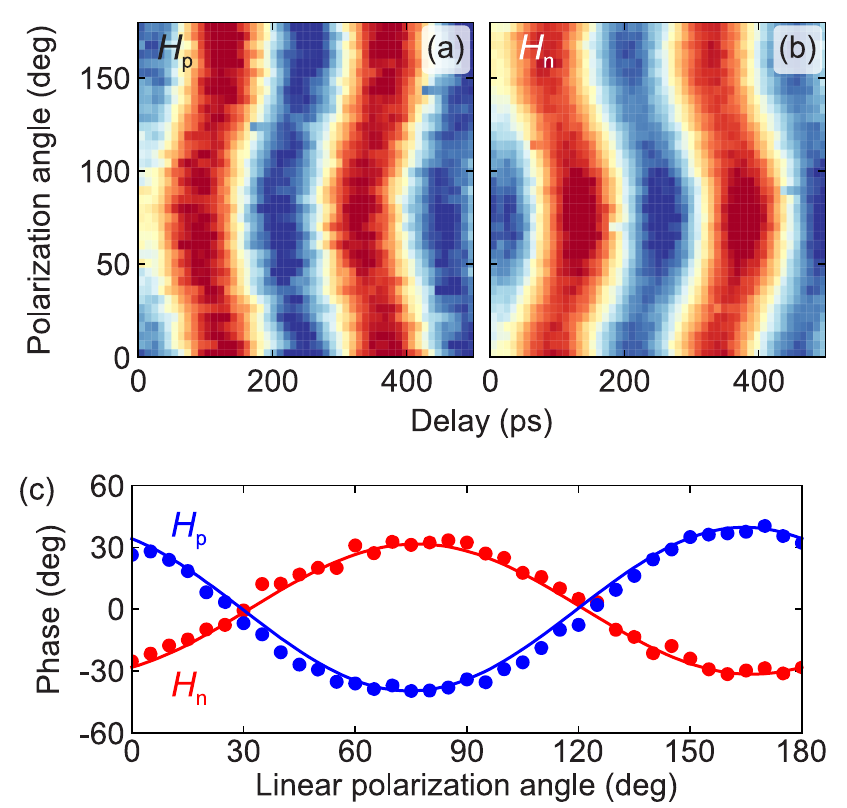}
	\caption{Dynamic contribution to the (001) Bragg peak of the conical phase of \cso\ measured at the Cu $L_3$ edge with linearly polarized light as a function of the linear polarization angle and the delay between RF excitation and x-ray probe pulse.
		The applied field was oriented in-plane, and (a) parallel ($H_p$) and (b) antiparallel ($H_n$) to the projection of the incident beam onto the sample plane.
		The sample was excited with RF at 4\,GHz in an applied field of 43\,mT at 25\,K, i.e., exciting the conical $+Q$ resonance.
		(c) Phase of the dynamic contribution to the (001) Bragg peak intensity as a function of the linear polarization angle.
	}
	\label{laa_delay}
\end{figure}

The VNA-FMR and XFMR measurements (with circularly polarized x-rays) allow us to identify the resonance conditions, and also obtain insight into the phase changes taking place over those resonances. 
However, further insight into the time evolution of the magnetic structure can be obtained by measuring the (001) Bragg diffraction of linearly polarized x-rays as a function of the linear polarization angle. 
The linearly polarized light introduces a mechanism to provide directional contrast into the measurements of the magnetic configuration of the system. 

We have shown earlier that the $+Q$ resonance mode in the conical phase in our \cso sample occurs at 4\,GHz in an in-plane bias field of 43\,mT at 25\,K.
With the system excited at this resonance, Fig.\ \ref{laa_delay}(a,b) shows the how the magnetic contrast during this dynamic process varies as a function of both the linear polarization angle and the time delay.

Similar to the measurements as a function of field in Fig.\ \ref{field_delay_amp_phase}, the dynamic signal has a sinusoidal dependence on the delay with a period of 250\,ps, originating from the 4\,GHz RF excitation frequency.
This dynamic signal shifts in phase as a function of the incident linear polarization angle.
The particular phase of $\sim$$30^\circ$ that is measured for a linear polarization angle of $0^\circ$ is given by the probed position within the resonance curve at 43\,mT [see Fig.\ \ref{field_delay_amp_phase}(b,c) for comparison].
Further, the reversal of the applied field direction reverses the direction of the phase shift as a function of the incident linear polarization angle, as shown in Fig.\ \ref{laa_delay}(a,b) for the projection of the field along the beam direction applied parallel and antiparallel, respectively. 
%
Sinusoidal fits to the delay scan data were used to extract the phase of the dynamic signal for both positive ($H_\mathrm{p}$) and negative field ($H_\mathrm{n}$), shown in Fig.\ \ref{laa_delay}(c). 
The phase of the dynamic signals in Fig.\ \ref{laa_delay}(c) varies sinusoidally with the linear polarization angle. 
Most importantly, the phase is inverted when the direction of the magnetic field is reversed. 
This phase inversion can be interpreted as an inversion of the polarization angle-dependent contrast, 
which in resonant scattering probes the contribution of the spins along and perpendicular to the polarization direction.

\begin{figure}[h!]
	\centering
	\includegraphics[width=86mm]{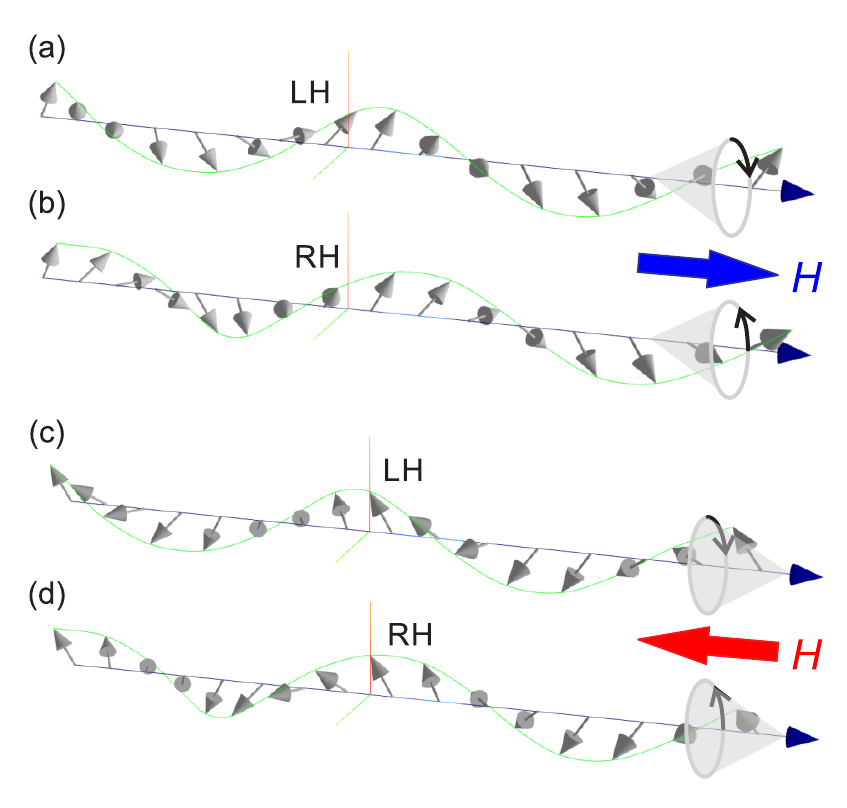}
	\caption{Illustration of equilibrium conical spin structures for applied fields along the helical propagation axis (a,b), and opposite to it (c,d), respectively.
		The sign of the Dzyaloshinskii–Moriya interaction (DMI), which is specific to the two enantiomers, results in (a,c) left-handed (LH) and (b,d) right-handed (RH) conical screws, respectively.
		See Supplemental Material for a closer look at the LH conical screw in a positive applied field shown in (a) \cite{suppl}.
	}
	\label{fig:static}
\end{figure}

In general, when the field direction is reversed and assuming that the spins simply follow the applied field, the conical spins change their equilibrium orientation, as illustrated in Fig.\ \ref{fig:static}.
Note that while the spins follow the magnetic field direction, the conical helix will remain either right- and left-handed, independent of the field direction.
As the rotation sense of the dynamical precession is determined by the direction of the applied field, which is counterclockwise for a free electron spin, the rotation sense reverses (in the reference frame of the sample) if the field direction is reversed.
As stated above, the conical dynamic mode which has the same (counterclockwise) rotation sense as a free spin, and therefore the field-polarized mode, is the $+Q$ modes.
Both modes are seamlessly connected at the transition field $H_{c2}$ (see upper row in Fig.\ \ref{vnafmr}).
Note that this behavior, i.e., the helicity of the modes, is unaffected by the chirality of the crystal \cite{Garst2017}.

\begin{figure*}[tbh]
	\centering
	\includegraphics[width=180mm]{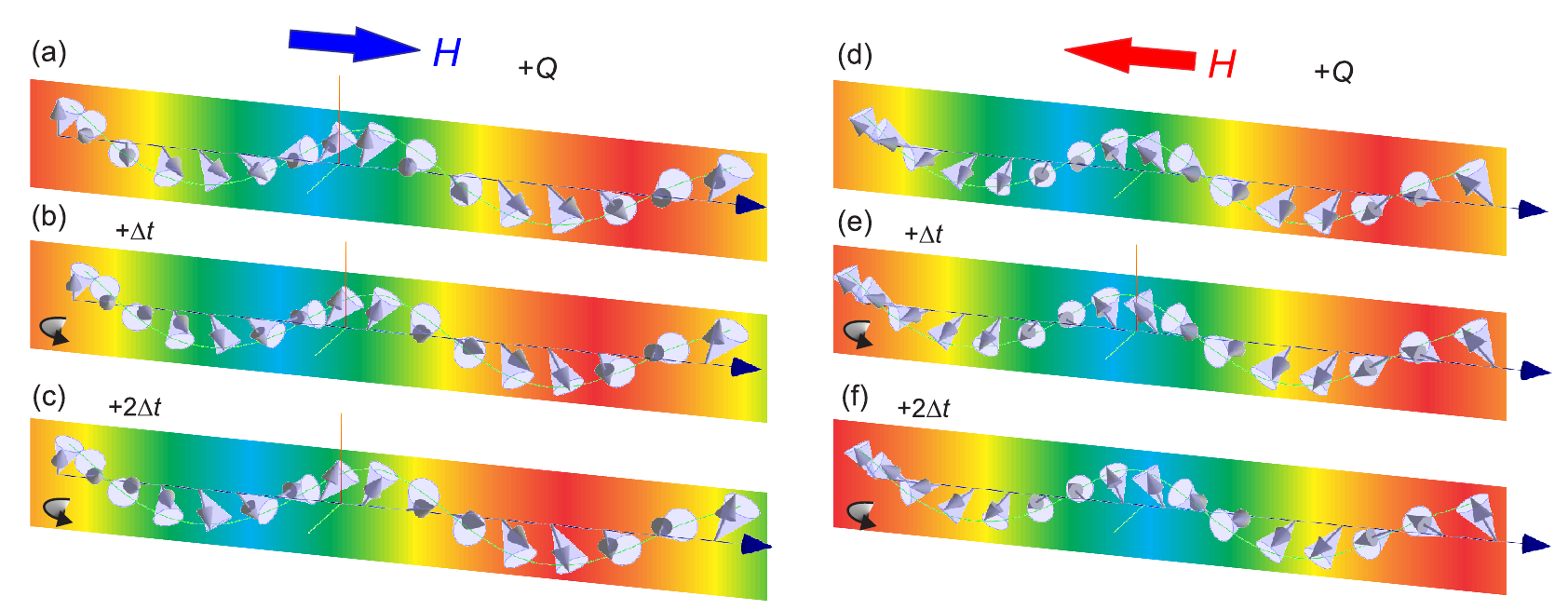}
	\caption{Illustration of the dynamic conical spin structures (for $+Q$) in applied fields along the helical propagation axis (a-c), and opposite to it (d-f), respectively.
		The panels show sequential snapshots at $t=0$, $\Delta t$, and $2 \Delta t$.
		The spins precess about the local effective field, describing a cone. 
		The axes of the cones coincide with the directions of the spins in the equilibrium conical structures (green lines) shown in Fig.\ \ref{fig:static}.
		A color scale has been superimposed as the background to indicate a compression wave where spins point closer (red) and further apart (blue) from one another. This compression wave propagates as a function of time. 
		The propagation of the compression wave is in opposite directions upon inversion of the applied field.
		See Supplemental Material for videos of both dynamical scenarios \cite{suppl}.
	}
	\label{fig:dynamic}
\end{figure*}

In our experiments, we selected the $+Q$ mode via the RF frequency at a given applied field.
Figure \ref{fig:dynamic} shows an illustration of the dynamic behavior of the $+Q$ mode upon field reversal.
The panels show subsequent snapshots of the precession of the spins about the local effective field.
For a positive applied field (along the helical propagation axis), a compression wave propagates to the right, in the direction of the applied field.
For a negative applied field, the propagation direction inverts as well.
This model is consistent with our experimental results (Fig.\ \ref{laa_delay}).

\section{Conclusions}

In summary, we explored the dynamic phase space of the resonance modes in the conical and field-polarized states in \cso\ as a function of field, temperature, and RF excitation frequency. 
First, we identified the resonances and measured their field and temperature dependence using VNA-FMR.
Taking into account the shape (and demagnetization factors) of the \cso\ sample \cite{Schwarze_NMat_2015}, the observed modes could be reproduced using the established models \cite{Garst2017}.
Employing the same setup, synchrotron-based DFMR allows for a deeper insight into the dynamics of the identified resonance modes.
We observed an inversion of the phase of the dynamic signal at 4\,GHz across the resonances in the field-polarized and conical state, owing to the way the resonances are traversed.
Furthermore, linearly polarized incident x-rays provide additional contrast, projecting out the specific orientation of the magnetization along the incident beam propagation direction, giving access to the time-dependent orientation of the magnetic spins during precession. 
We find that upon reversing the applied magnetic field direction, the direction of the propagating compression wave is inverted.
In general, being able to measure the time-dependent orientation of the magnetization allows for the exploration of the dynamics of chiral systems, which has so far only been accessible via RF absorption measurements in VNA-FMR.
The technique of DFMR therefore provides a unique view into the magnetization dynamics of complex topologically structured spin systems.


\section{Acknowledgments}
We thank Prof.\ Markus Garst from KIT in Karlsruhe for fruitful discussion about the dynamical behavior of magnetization patterns in \cso.
We acknowledge the Diamond Light Source (Oxfordshire, UK) for the provision of synchrotron radiation beamtime at beamline I10 under proposals SI16043 and SI17612.
This research was funded in part by the Engineering and Physical Sciences Research Council (UK) through grant EP/N032128/1.
For the purpose of Open Access, the author has applied a CC BY public copyright license to any Author Accepted Manuscript version arising from this submission.



\begin{thebibliography}{39}%
	\makeatletter
	\providecommand \@ifxundefined [1]{%
		\@ifx{#1\undefined}
	}%
	\providecommand \@ifnum [1]{%
		\ifnum #1\expandafter \@firstoftwo
		\else \expandafter \@secondoftwo
		\fi
	}%
	\providecommand \@ifx [1]{%
		\ifx #1\expandafter \@firstoftwo
		\else \expandafter \@secondoftwo
		\fi
	}%
	\providecommand \natexlab [1]{#1}%
	\providecommand \enquote  [1]{``#1''}%
	\providecommand \bibnamefont  [1]{#1}%
	\providecommand \bibfnamefont [1]{#1}%
	\providecommand \citenamefont [1]{#1}%
	\providecommand \href@noop [0]{\@secondoftwo}%
	\providecommand \href [0]{\begingroup \@sanitize@url \@href}%
	\providecommand \@href[1]{\@@startlink{#1}\@@href}%
	\providecommand \@@href[1]{\endgroup#1\@@endlink}%
	\providecommand \@sanitize@url [0]{\catcode `\\12\catcode `\$12\catcode
		`\&12\catcode `\#12\catcode `\^12\catcode `\_12\catcode `\%12\relax}%
	\providecommand \@@startlink[1]{}%
	\providecommand \@@endlink[0]{}%
	\providecommand \url  [0]{\begingroup\@sanitize@url \@url }%
	\providecommand \@url [1]{\endgroup\@href {#1}{\urlprefix }}%
	\providecommand \urlprefix  [0]{URL }%
	\providecommand \Eprint [0]{\href }%
	\providecommand \doibase [0]{https://doi.org/}%
	\providecommand \selectlanguage [0]{\@gobble}%
	\providecommand \bibinfo  [0]{\@secondoftwo}%
	\providecommand \bibfield  [0]{\@secondoftwo}%
	\providecommand \translation [1]{[#1]}%
	\providecommand \BibitemOpen [0]{}%
	\providecommand \bibitemStop [0]{}%
	\providecommand \bibitemNoStop [0]{.\EOS\space}%
	\providecommand \EOS [0]{\spacefactor3000\relax}%
	\providecommand \BibitemShut  [1]{\csname bibitem#1\endcsname}%
	\let\auto@bib@innerbib\@empty
	\bibitem [{\citenamefont {Onose}\ \emph {et~al.}(2012)\citenamefont {Onose},
		\citenamefont {Okamura}, \citenamefont {Seki}, \citenamefont {Ishiwata},\
		and\ \citenamefont {Tokura}}]{Onose_PRL_2012}%
	\BibitemOpen
	\bibfield  {author} {\bibinfo {author} {\bibfnamefont {Y.}~\bibnamefont
			{Onose}}, \bibinfo {author} {\bibfnamefont {Y.}~\bibnamefont {Okamura}},
		\bibinfo {author} {\bibfnamefont {S.}~\bibnamefont {Seki}}, \bibinfo {author}
		{\bibfnamefont {S.}~\bibnamefont {Ishiwata}},\ and\ \bibinfo {author}
		{\bibfnamefont {Y.}~\bibnamefont {Tokura}},\ }\bibfield  {title} {\bibinfo
		{title} {{Observation of magnetic excitations of skyrmion crystal in a
				helimagnetic insulator Cu 2OSeO 3}},\ }\href
	{https://doi.org/10.1103/PhysRevLett.109.037603} {\bibfield  {journal}
		{\bibinfo  {journal} {Phys. Rev. Lett.}\ }\textbf {\bibinfo {volume} {109}},\
		\bibinfo {pages} {037603} (\bibinfo {year} {2012})}\BibitemShut {NoStop}%
	\bibitem [{\citenamefont {Nagaosa}\ and\ \citenamefont
		{Tokura}(2013)}]{NatNano_8_899}%
	\BibitemOpen
	\bibfield  {author} {\bibinfo {author} {\bibfnamefont {N.}~\bibnamefont
			{Nagaosa}}\ and\ \bibinfo {author} {\bibfnamefont {Y.}~\bibnamefont
			{Tokura}},\ }\bibfield  {title} {\bibinfo {title} {{Topological properties
				and dynamics of magnetic skyrmions}},\ }\href
	{https://doi.org/10.1038/nnano.2013.243} {\bibfield  {journal} {\bibinfo
			{journal} {Nat. Nanotechnol.}\ }\textbf {\bibinfo {volume} {8}},\ \bibinfo
		{pages} {899} (\bibinfo {year} {2013})}\BibitemShut {NoStop}%
	\bibitem [{\citenamefont {Cochran}\ \emph {et~al.}(1986)\citenamefont
		{Cochran}, \citenamefont {Heinrich},\ and\ \citenamefont
		{Arrott}}]{Cochran1986}%
	\BibitemOpen
	\bibfield  {author} {\bibinfo {author} {\bibfnamefont {J.~F.}\ \bibnamefont
			{Cochran}}, \bibinfo {author} {\bibfnamefont {B.}~\bibnamefont {Heinrich}},\
		and\ \bibinfo {author} {\bibfnamefont {A.~S.}\ \bibnamefont {Arrott}},\
	}\bibfield  {title} {\bibinfo {title} {{Ferromagnetic-resonance in a system
				composed of a ferromagnetic substrate and an exchange-coupled thin
				ferromagnetic overlayer}},\ }\href {https://doi.org/10.1103/PhysRevB.34.7788}
	{\bibfield  {journal} {\bibinfo  {journal} {Phys. Rev. B}\ }\textbf {\bibinfo
			{volume} {34}},\ \bibinfo {pages} {7788} (\bibinfo {year}
		{1986})}\BibitemShut {NoStop}%
	\bibitem [{\citenamefont {Metaxas}\ \emph {et~al.}(2010)\citenamefont
		{Metaxas}, \citenamefont {Stamps}, \citenamefont {Jamet}, \citenamefont
		{Ferre}, \citenamefont {Baltz}, \citenamefont {Rodmacq},\ and\ \citenamefont
		{Politi}}]{Metaxas2010}%
	\BibitemOpen
	\bibfield  {author} {\bibinfo {author} {\bibfnamefont {P.~J.}\ \bibnamefont
			{Metaxas}}, \bibinfo {author} {\bibfnamefont {R.~L.}\ \bibnamefont {Stamps}},
		\bibinfo {author} {\bibfnamefont {J.-P.}\ \bibnamefont {Jamet}}, \bibinfo
		{author} {\bibfnamefont {J.}~\bibnamefont {Ferre}}, \bibinfo {author}
		{\bibfnamefont {V.}~\bibnamefont {Baltz}}, \bibinfo {author} {\bibfnamefont
			{B.}~\bibnamefont {Rodmacq}},\ and\ \bibinfo {author} {\bibfnamefont
			{P.}~\bibnamefont {Politi}},\ }\bibfield  {title} {\bibinfo {title} {{Dynamic
				binding of driven interfaces in coupled ultrathin ferromagnetic layers}},\
	}\href {https://doi.org/{10.1103/PhysRevLett.104.237206}} {\bibfield
		{journal} {\bibinfo  {journal} {{Phys. Rev. Lett.}}\ }\textbf {\bibinfo
			{volume} {{104}}},\ \bibinfo {pages} {237206} (\bibinfo {year}
		{{2010}})}\BibitemShut {NoStop}%
	\bibitem [{\citenamefont {Magaraggia}\ \emph {et~al.}(2011)\citenamefont
		{Magaraggia}, \citenamefont {Kennewell}, \citenamefont {Kostylev},
		\citenamefont {Stamps}, \citenamefont {Ali}, \citenamefont {Greig},
		\citenamefont {Hickey},\ and\ \citenamefont {Marrows}}]{Magaraggia2011}%
	\BibitemOpen
	\bibfield  {author} {\bibinfo {author} {\bibfnamefont {R.}~\bibnamefont
			{Magaraggia}}, \bibinfo {author} {\bibfnamefont {K.}~\bibnamefont
			{Kennewell}}, \bibinfo {author} {\bibfnamefont {M.}~\bibnamefont {Kostylev}},
		\bibinfo {author} {\bibfnamefont {R.~L.}\ \bibnamefont {Stamps}}, \bibinfo
		{author} {\bibfnamefont {M.}~\bibnamefont {Ali}}, \bibinfo {author}
		{\bibfnamefont {D.}~\bibnamefont {Greig}}, \bibinfo {author} {\bibfnamefont
			{B.~J.}\ \bibnamefont {Hickey}},\ and\ \bibinfo {author} {\bibfnamefont
			{C.~H.}\ \bibnamefont {Marrows}},\ }\bibfield  {title} {\bibinfo {title}
		{{Exchange anisotropy pinning of a standing spin-wave mode}},\ }\href
	{https://doi.org/{10.1103/PhysRevB.83.054405}} {\bibfield  {journal}
		{\bibinfo  {journal} {{Phys. Rev. B}}\ }\textbf {\bibinfo {volume} {{83}}},\
		\bibinfo {pages} {054405} (\bibinfo {year} {{2011}})}\BibitemShut {NoStop}%
	\bibitem [{\citenamefont {Kaiser}\ \emph {et~al.}(2011)\citenamefont {Kaiser},
		\citenamefont {Schoeppner}, \citenamefont {Roemer}, \citenamefont {Hassel},
		\citenamefont {Wiemann}, \citenamefont {Cramm}, \citenamefont {Nickel},
		\citenamefont {Grychtol}, \citenamefont {Tieg}, \citenamefont {Lindner},\
		and\ \citenamefont {Schneider}}]{Kaiser2011}%
	\BibitemOpen
	\bibfield  {author} {\bibinfo {author} {\bibfnamefont {A.~M.}\ \bibnamefont
			{Kaiser}}, \bibinfo {author} {\bibfnamefont {C.}~\bibnamefont {Schoeppner}},
		\bibinfo {author} {\bibfnamefont {F.~M.}\ \bibnamefont {Roemer}}, \bibinfo
		{author} {\bibfnamefont {C.}~\bibnamefont {Hassel}}, \bibinfo {author}
		{\bibfnamefont {C.}~\bibnamefont {Wiemann}}, \bibinfo {author} {\bibfnamefont
			{S.}~\bibnamefont {Cramm}}, \bibinfo {author} {\bibfnamefont
			{F.}~\bibnamefont {Nickel}}, \bibinfo {author} {\bibfnamefont
			{P.}~\bibnamefont {Grychtol}}, \bibinfo {author} {\bibfnamefont
			{C.}~\bibnamefont {Tieg}}, \bibinfo {author} {\bibfnamefont {J.}~\bibnamefont
			{Lindner}},\ and\ \bibinfo {author} {\bibfnamefont {C.~M.}\ \bibnamefont
			{Schneider}},\ }\bibfield  {title} {\bibinfo {title} {{Nano and picosecond
				magnetization dynamics of weakly coupled CoFe/Cr/NiFe trilayers studied by a
				multitechnique approach}},\ }\href
	{https://doi.org/{10.1103/PhysRevB.84.134406}} {\bibfield  {journal}
		{\bibinfo  {journal} {{Phys. Rev. B}}\ }\textbf {\bibinfo {volume} {{84}}},\
		\bibinfo {pages} {134406} (\bibinfo {year} {{2011}})}\BibitemShut {NoStop}%
	\bibitem [{\citenamefont {Schwarze}\ \emph {et~al.}(2015)\citenamefont
		{Schwarze}, \citenamefont {Waizner}, \citenamefont {Garst}, \citenamefont
		{Bauer}, \citenamefont {Stasinopoulos}, \citenamefont {Berger}, \citenamefont
		{Pfleiderer},\ and\ \citenamefont {Grundler}}]{Schwarze_NMat_2015}%
	\BibitemOpen
	\bibfield  {author} {\bibinfo {author} {\bibfnamefont {T.}~\bibnamefont
			{Schwarze}}, \bibinfo {author} {\bibfnamefont {J.}~\bibnamefont {Waizner}},
		\bibinfo {author} {\bibfnamefont {M.}~\bibnamefont {Garst}}, \bibinfo
		{author} {\bibfnamefont {A.}~\bibnamefont {Bauer}}, \bibinfo {author}
		{\bibfnamefont {I.}~\bibnamefont {Stasinopoulos}}, \bibinfo {author}
		{\bibfnamefont {H.}~\bibnamefont {Berger}}, \bibinfo {author} {\bibfnamefont
			{C.}~\bibnamefont {Pfleiderer}},\ and\ \bibinfo {author} {\bibfnamefont
			{D.}~\bibnamefont {Grundler}},\ }\bibfield  {title} {\bibinfo {title}
		{{Universal helimagnon and skyrmion excitations in metallic, semiconducting
				and insulating chiral magnets}},\ }\href {https://doi.org/10.1038/nmat4223}
	{\bibfield  {journal} {\bibinfo  {journal} {Nat. Mater.}\ }\textbf {\bibinfo
			{volume} {14}},\ \bibinfo {pages} {478} (\bibinfo {year} {2015})}\BibitemShut
	{NoStop}%
	\bibitem [{\citenamefont {Stasinopoulos}\ \emph {et~al.}(2017)\citenamefont
		{Stasinopoulos}, \citenamefont {Weichselbaumer}, \citenamefont {Bauer},
		\citenamefont {Waizner}, \citenamefont {Berger}, \citenamefont {Garst},
		\citenamefont {Pfleiderer},\ and\ \citenamefont
		{Grundler}}]{Stasinopoulos_SciRep_2017}%
	\BibitemOpen
	\bibfield  {author} {\bibinfo {author} {\bibfnamefont {I.}~\bibnamefont
			{Stasinopoulos}}, \bibinfo {author} {\bibfnamefont {S.}~\bibnamefont
			{Weichselbaumer}}, \bibinfo {author} {\bibfnamefont {A.}~\bibnamefont
			{Bauer}}, \bibinfo {author} {\bibfnamefont {J.}~\bibnamefont {Waizner}},
		\bibinfo {author} {\bibfnamefont {H.}~\bibnamefont {Berger}}, \bibinfo
		{author} {\bibfnamefont {M.}~\bibnamefont {Garst}}, \bibinfo {author}
		{\bibfnamefont {C.}~\bibnamefont {Pfleiderer}},\ and\ \bibinfo {author}
		{\bibfnamefont {D.}~\bibnamefont {Grundler}},\ }\bibfield  {title} {\bibinfo
		{title} {{Linearly polarized GHz magnetization dynamics of spin helix modes
				in the ferrimagnetic insulator Cu$_2$OSeO$_3$}},\ }\href
	{https://doi.org/10.1038/s41598-017-07020-2} {\bibfield  {journal} {\bibinfo
			{journal} {Sci. Rep.}\ }\textbf {\bibinfo {volume} {7}},\ \bibinfo {pages}
		{7037} (\bibinfo {year} {2017})}\BibitemShut {NoStop}%
	\bibitem [{\citenamefont {Weiler}\ \emph {et~al.}(2017)\citenamefont {Weiler},
		\citenamefont {Aqeel}, \citenamefont {Mostovoy}, \citenamefont {Leonov},
		\citenamefont {Gepr{\"{a}}gs}, \citenamefont {Gross}, \citenamefont {Huebl},
		\citenamefont {Palstra},\ and\ \citenamefont
		{Goennenwein}}]{Weiler_PRL_2017}%
	\BibitemOpen
	\bibfield  {author} {\bibinfo {author} {\bibfnamefont {M.}~\bibnamefont
			{Weiler}}, \bibinfo {author} {\bibfnamefont {A.}~\bibnamefont {Aqeel}},
		\bibinfo {author} {\bibfnamefont {M.}~\bibnamefont {Mostovoy}}, \bibinfo
		{author} {\bibfnamefont {A.}~\bibnamefont {Leonov}}, \bibinfo {author}
		{\bibfnamefont {S.}~\bibnamefont {Gepr{\"{a}}gs}}, \bibinfo {author}
		{\bibfnamefont {R.}~\bibnamefont {Gross}}, \bibinfo {author} {\bibfnamefont
			{H.}~\bibnamefont {Huebl}}, \bibinfo {author} {\bibfnamefont {T.~T.~M.}\
			\bibnamefont {Palstra}},\ and\ \bibinfo {author} {\bibfnamefont {S.~T.~B.}\
			\bibnamefont {Goennenwein}},\ }\bibfield  {title} {\bibinfo {title}
		{{Helimagnon resonances in an intrinsic chiral magnonic crystal}},\ }\href
	{https://doi.org/10.1103/PhysRevLett.119.237204} {\bibfield  {journal}
		{\bibinfo  {journal} {Phys. Rev. Lett.}\ }\textbf {\bibinfo {volume} {119}},\
		\bibinfo {pages} {237204} (\bibinfo {year} {2017})}\BibitemShut {NoStop}%
	\bibitem [{\citenamefont {Garst}\ \emph {et~al.}(2017)\citenamefont {Garst},
		\citenamefont {Waizner},\ and\ \citenamefont {Grundler}}]{Garst2017}%
	\BibitemOpen
	\bibfield  {author} {\bibinfo {author} {\bibfnamefont {M.}~\bibnamefont
			{Garst}}, \bibinfo {author} {\bibfnamefont {J.}~\bibnamefont {Waizner}},\
		and\ \bibinfo {author} {\bibfnamefont {D.}~\bibnamefont {Grundler}},\
	}\bibfield  {title} {\bibinfo {title} {{Collective spin excitations of
				helices and magnetic skyrmions: review and perspectives of magnonics in
				non-centrosymmetrric magnets}},\ }\href@noop {} {\bibfield  {journal}
		{\bibinfo  {journal} {J. Phys. D: Appl. Phys}\ }\textbf {\bibinfo {volume}
			{50}},\ \bibinfo {pages} {293002} (\bibinfo {year} {2017})}\BibitemShut
	{NoStop}%
	\bibitem [{\citenamefont {Zhang}\ \emph
		{et~al.}(2016{\natexlab{a}})\citenamefont {Zhang}, \citenamefont {Bauer},
		\citenamefont {Berger}, \citenamefont {Pfleiderer}, \citenamefont {van~der
			Laan},\ and\ \citenamefont {Hesjedal}}]{PRB_93_214420}%
	\BibitemOpen
	\bibfield  {author} {\bibinfo {author} {\bibfnamefont {S.~L.}\ \bibnamefont
			{Zhang}}, \bibinfo {author} {\bibfnamefont {A.}~\bibnamefont {Bauer}},
		\bibinfo {author} {\bibfnamefont {H.}~\bibnamefont {Berger}}, \bibinfo
		{author} {\bibfnamefont {C.}~\bibnamefont {Pfleiderer}}, \bibinfo {author}
		{\bibfnamefont {G.}~\bibnamefont {van~der Laan}},\ and\ \bibinfo {author}
		{\bibfnamefont {T.}~\bibnamefont {Hesjedal}},\ }\bibfield  {title} {\bibinfo
		{title} {{Resonant elastic x-ray scattering from the skyrmion lattice in
				Cu$_2$OSeO$_3$}},\ }\href {https://doi.org/10.1103/PhysRevB.93.214420}
	{\bibfield  {journal} {\bibinfo  {journal} {Phys. Rev. B}\ }\textbf {\bibinfo
			{volume} {93}},\ \bibinfo {pages} {214420} (\bibinfo {year}
		{2016}{\natexlab{a}})}\BibitemShut {NoStop}%
	\bibitem [{\citenamefont {Zhang}\ \emph
		{et~al.}(2016{\natexlab{b}})\citenamefont {Zhang}, \citenamefont {Bauer},
		\citenamefont {Burn}, \citenamefont {Milde}, \citenamefont {Neuber},
		\citenamefont {Eng}, \citenamefont {Berger}, \citenamefont {Pfleiderer},
		\citenamefont {van~der Laan},\ and\ \citenamefont
		{Hesjedal}}]{NanoLett_16_3285}%
	\BibitemOpen
	\bibfield  {author} {\bibinfo {author} {\bibfnamefont {S.~L.}\ \bibnamefont
			{Zhang}}, \bibinfo {author} {\bibfnamefont {A.}~\bibnamefont {Bauer}},
		\bibinfo {author} {\bibfnamefont {D.~M.}\ \bibnamefont {Burn}}, \bibinfo
		{author} {\bibfnamefont {P.}~\bibnamefont {Milde}}, \bibinfo {author}
		{\bibfnamefont {E.}~\bibnamefont {Neuber}}, \bibinfo {author} {\bibfnamefont
			{L.~M.}\ \bibnamefont {Eng}}, \bibinfo {author} {\bibfnamefont
			{H.}~\bibnamefont {Berger}}, \bibinfo {author} {\bibfnamefont
			{C.}~\bibnamefont {Pfleiderer}}, \bibinfo {author} {\bibfnamefont
			{G.}~\bibnamefont {van~der Laan}},\ and\ \bibinfo {author} {\bibfnamefont
			{T.}~\bibnamefont {Hesjedal}},\ }\bibfield  {title} {\bibinfo {title}
		{{Multidomain Skyrmion Lattice State in Cu$_2$OSeO$_3$}},\ }\href
	{https://doi.org/10.1021/acs.nanolett.6b00845} {\bibfield  {journal}
		{\bibinfo  {journal} {Nano Lett.}\ }\textbf {\bibinfo {volume} {16}},\
		\bibinfo {pages} {3285} (\bibinfo {year} {2016}{\natexlab{b}})}\BibitemShut
	{NoStop}%
	\bibitem [{\citenamefont {Zhang}\ \emph
		{et~al.}(2016{\natexlab{c}})\citenamefont {Zhang}, \citenamefont {Bauer},
		\citenamefont {Berger}, \citenamefont {Pfleiderer}, \citenamefont {van~der
			Laan},\ and\ \citenamefont {Hesjedal}}]{CSO-APL_2016}%
	\BibitemOpen
	\bibfield  {author} {\bibinfo {author} {\bibfnamefont {S.~L.}\ \bibnamefont
			{Zhang}}, \bibinfo {author} {\bibfnamefont {A.}~\bibnamefont {Bauer}},
		\bibinfo {author} {\bibfnamefont {H.}~\bibnamefont {Berger}}, \bibinfo
		{author} {\bibfnamefont {C.}~\bibnamefont {Pfleiderer}}, \bibinfo {author}
		{\bibfnamefont {G.}~\bibnamefont {van~der Laan}},\ and\ \bibinfo {author}
		{\bibfnamefont {T.}~\bibnamefont {Hesjedal}},\ }\bibfield  {title} {\bibinfo
		{title} {{Imaging and manipulation of skyrmion lattice domains in
				Cu$_2$OSeO$_3$}},\ }\href {https://doi.org/10.1063/1.4967499} {\bibfield
		{journal} {\bibinfo  {journal} {Appl. Phys. Lett.}\ }\textbf {\bibinfo
			{volume} {109}},\ \bibinfo {pages} {192406} (\bibinfo {year}
		{2016}{\natexlab{c}})}\BibitemShut {NoStop}%
	\bibitem [{\citenamefont {Zhang}\ \emph
		{et~al.}(2017{\natexlab{a}})\citenamefont {Zhang}, \citenamefont {van~der
			Laan},\ and\ \citenamefont {Hesjedal}}]{NatComms_8_14619}%
	\BibitemOpen
	\bibfield  {author} {\bibinfo {author} {\bibfnamefont {S.~L.}\ \bibnamefont
			{Zhang}}, \bibinfo {author} {\bibfnamefont {G.}~\bibnamefont {van~der
				Laan}},\ and\ \bibinfo {author} {\bibfnamefont {T.}~\bibnamefont
			{Hesjedal}},\ }\bibfield  {title} {\bibinfo {title} {{Direct experimental
				determination of the topological winding number of skyrmions in
				Cu$_2$OSeO$_3$}},\ }\href {https://doi.org/10.1038/ncomms14619} {\bibfield
		{journal} {\bibinfo  {journal} {Nat. Commun.}\ }\textbf {\bibinfo {volume}
			{8}},\ \bibinfo {pages} {14619} (\bibinfo {year}
		{2017}{\natexlab{a}})}\BibitemShut {NoStop}%
	\bibitem [{\citenamefont {Ukleev}\ \emph {et~al.}(2022)\citenamefont {Ukleev},
		\citenamefont {Luo}, \citenamefont {Abrudan}, \citenamefont {Aqeel},
		\citenamefont {Back},\ and\ \citenamefont {Radu}}]{Ukleev_2022}%
	\BibitemOpen
	\bibfield  {author} {\bibinfo {author} {\bibfnamefont {V.}~\bibnamefont
			{Ukleev}}, \bibinfo {author} {\bibfnamefont {C.}~\bibnamefont {Luo}},
		\bibinfo {author} {\bibfnamefont {R.}~\bibnamefont {Abrudan}}, \bibinfo
		{author} {\bibfnamefont {A.}~\bibnamefont {Aqeel}}, \bibinfo {author}
		{\bibfnamefont {C.~H.}\ \bibnamefont {Back}},\ and\ \bibinfo {author}
		{\bibfnamefont {F.}~\bibnamefont {Radu}},\ }\bibfield  {title} {\bibinfo
		{title} {{Chiral surface spin textures in Cu$_2$OSeO$_3$ unveiled by soft
				X-ray scattering in specular reflection geometry}},\ }\href
	{https://doi.org/10.1080/14686996.2022.2131466} {\bibfield  {journal}
		{\bibinfo  {journal} {Sci. Technol. Adv. Mater.}\ }\textbf {\bibinfo {volume}
			{23}},\ \bibinfo {pages} {682} (\bibinfo {year} {2022})}\BibitemShut
	{NoStop}%
	\bibitem [{\citenamefont {D{\"u}rr}\ \emph {et~al.}(1999)\citenamefont
		{D{\"u}rr}, \citenamefont {Dudzik}, \citenamefont {Dhesi}, \citenamefont
		{Goedkoop}, \citenamefont {van~der Laan}, \citenamefont {Belakhovsky},
		\citenamefont {Mocuta}, \citenamefont {Marty},\ and\ \citenamefont
		{Samson}}]{Duerr1999}%
	\BibitemOpen
	\bibfield  {author} {\bibinfo {author} {\bibfnamefont {H.~A.}\ \bibnamefont
			{D{\"u}rr}}, \bibinfo {author} {\bibfnamefont {E.}~\bibnamefont {Dudzik}},
		\bibinfo {author} {\bibfnamefont {S.~S.}\ \bibnamefont {Dhesi}}, \bibinfo
		{author} {\bibfnamefont {J.~B.}\ \bibnamefont {Goedkoop}}, \bibinfo {author}
		{\bibfnamefont {G.}~\bibnamefont {van~der Laan}}, \bibinfo {author}
		{\bibfnamefont {M.}~\bibnamefont {Belakhovsky}}, \bibinfo {author}
		{\bibfnamefont {C.}~\bibnamefont {Mocuta}}, \bibinfo {author} {\bibfnamefont
			{A.}~\bibnamefont {Marty}},\ and\ \bibinfo {author} {\bibfnamefont
			{Y.}~\bibnamefont {Samson}},\ }\bibfield  {title} {\bibinfo {title} {{Chiral
				Magnetic Domain Structures in Ultrathin FePd Films}},\ }\href
	{https://doi.org/10.1126/science.284.5423.2166} {\bibfield  {journal}
		{\bibinfo  {journal} {Science}\ }\textbf {\bibinfo {volume} {284}},\ \bibinfo
		{pages} {2166} (\bibinfo {year} {1999})}\BibitemShut {NoStop}%
	\bibitem [{\citenamefont {Zhang}\ \emph
		{et~al.}(2017{\natexlab{b}})\citenamefont {Zhang}, \citenamefont {van~der
			Laan},\ and\ \citenamefont {Hesjedal}}]{Zhang_PRB_2017}%
	\BibitemOpen
	\bibfield  {author} {\bibinfo {author} {\bibfnamefont {S.~L.}\ \bibnamefont
			{Zhang}}, \bibinfo {author} {\bibfnamefont {G.}~\bibnamefont {van~der
				Laan}},\ and\ \bibinfo {author} {\bibfnamefont {T.}~\bibnamefont
			{Hesjedal}},\ }\bibfield  {title} {\bibinfo {title} {{Direct experimental
				determination of spiral spin structures via the dichroism extinction effect
				in resonant elastic soft x-ray scattering}},\ }\href
	{https://doi.org/10.1103/PhysRevB.96.094401} {\bibfield  {journal} {\bibinfo
			{journal} {Phys. Rev. B}\ }\textbf {\bibinfo {volume} {96}},\ \bibinfo
		{pages} {094401} (\bibinfo {year} {2017}{\natexlab{b}})}\BibitemShut
	{NoStop}%
	\bibitem [{\citenamefont {Chauleau}\ \emph {et~al.}(2018)\citenamefont
		{Chauleau}, \citenamefont {Legrand}, \citenamefont {Reyren}, \citenamefont
		{Maccariello}, \citenamefont {Collin}, \citenamefont {Popescu}, \citenamefont
		{Bouzehouane}, \citenamefont {Cros}, \citenamefont {Jaouen},\ and\
		\citenamefont {Fert}}]{PhysRevLett.120.037202}%
	\BibitemOpen
	\bibfield  {author} {\bibinfo {author} {\bibfnamefont {J.-Y.}\ \bibnamefont
			{Chauleau}}, \bibinfo {author} {\bibfnamefont {W.}~\bibnamefont {Legrand}},
		\bibinfo {author} {\bibfnamefont {N.}~\bibnamefont {Reyren}}, \bibinfo
		{author} {\bibfnamefont {D.}~\bibnamefont {Maccariello}}, \bibinfo {author}
		{\bibfnamefont {S.}~\bibnamefont {Collin}}, \bibinfo {author} {\bibfnamefont
			{H.}~\bibnamefont {Popescu}}, \bibinfo {author} {\bibfnamefont
			{K.}~\bibnamefont {Bouzehouane}}, \bibinfo {author} {\bibfnamefont
			{V.}~\bibnamefont {Cros}}, \bibinfo {author} {\bibfnamefont {N.}~\bibnamefont
			{Jaouen}},\ and\ \bibinfo {author} {\bibfnamefont {A.}~\bibnamefont {Fert}},\
	}\bibfield  {title} {\bibinfo {title} {Chirality in magnetic multilayers
			probed by the symmetry and the amplitude of dichroism in x-ray resonant
			magnetic scattering},\ }\href
	{https://doi.org/10.1103/PhysRevLett.120.037202} {\bibfield  {journal}
		{\bibinfo  {journal} {Phys. Rev. Lett.}\ }\textbf {\bibinfo {volume} {120}},\
		\bibinfo {pages} {037202} (\bibinfo {year} {2018})}\BibitemShut {NoStop}%
	\bibitem [{\citenamefont {Legrand}\ \emph {et~al.}(2018)\citenamefont
		{Legrand}, \citenamefont {Chauleau}, \citenamefont {Maccariello},
		\citenamefont {Reyren}, \citenamefont {Collin}, \citenamefont {Bouzehouane},
		\citenamefont {Jaouen}, \citenamefont {Cros},\ and\ \citenamefont
		{Fert}}]{Legrand2018}%
	\BibitemOpen
	\bibfield  {author} {\bibinfo {author} {\bibfnamefont {W.}~\bibnamefont
			{Legrand}}, \bibinfo {author} {\bibfnamefont {J.-Y.}\ \bibnamefont
			{Chauleau}}, \bibinfo {author} {\bibfnamefont {D.}~\bibnamefont
			{Maccariello}}, \bibinfo {author} {\bibfnamefont {N.}~\bibnamefont {Reyren}},
		\bibinfo {author} {\bibfnamefont {S.}~\bibnamefont {Collin}}, \bibinfo
		{author} {\bibfnamefont {K.}~\bibnamefont {Bouzehouane}}, \bibinfo {author}
		{\bibfnamefont {N.}~\bibnamefont {Jaouen}}, \bibinfo {author} {\bibfnamefont
			{V.}~\bibnamefont {Cros}},\ and\ \bibinfo {author} {\bibfnamefont
			{A.}~\bibnamefont {Fert}},\ }\bibfield  {title} {\bibinfo {title} {Hybrid
			chiral domain walls and skyrmions in magnetic multilayers},\ }\href
	{https://doi.org/10.1126/sciadv.aat0415} {\bibfield  {journal} {\bibinfo
			{journal} {Sci. Adv.}\ }\textbf {\bibinfo {volume} {4}},\ \bibinfo {pages}
		{eaat0415} (\bibinfo {year} {2018})}\BibitemShut {NoStop}%
	\bibitem [{\citenamefont {Li}\ \emph {et~al.}(2019)\citenamefont {Li},
		\citenamefont {Bykova}, \citenamefont {Zhang}, \citenamefont {Yu},
		\citenamefont {Tomasello}, \citenamefont {Carpentieri}, \citenamefont {Liu},
		\citenamefont {Guang}, \citenamefont {Gr{\"a}fe}, \citenamefont {Weigand},
		\citenamefont {Burn}, \citenamefont {van~der Laan}, \citenamefont {Hesjedal},
		\citenamefont {Yan}, \citenamefont {Feng}, \citenamefont {Wan}, \citenamefont
		{Wei}, \citenamefont {Wang}, \citenamefont {Zhang}, \citenamefont {Xu},
		\citenamefont {Guo}, \citenamefont {Wei}, \citenamefont {Finocchio},
		\citenamefont {Han},\ and\ \citenamefont {Sch{\"u}tz}}]{Li2019}%
	\BibitemOpen
	\bibfield  {author} {\bibinfo {author} {\bibfnamefont {W.}~\bibnamefont
			{Li}}, \bibinfo {author} {\bibfnamefont {I.}~\bibnamefont {Bykova}}, \bibinfo
		{author} {\bibfnamefont {S.}~\bibnamefont {Zhang}}, \bibinfo {author}
		{\bibfnamefont {G.}~\bibnamefont {Yu}}, \bibinfo {author} {\bibfnamefont
			{R.}~\bibnamefont {Tomasello}}, \bibinfo {author} {\bibfnamefont
			{M.}~\bibnamefont {Carpentieri}}, \bibinfo {author} {\bibfnamefont
			{Y.}~\bibnamefont {Liu}}, \bibinfo {author} {\bibfnamefont {Y.}~\bibnamefont
			{Guang}}, \bibinfo {author} {\bibfnamefont {J.}~\bibnamefont {Gr{\"a}fe}},
		\bibinfo {author} {\bibfnamefont {M.}~\bibnamefont {Weigand}}, \bibinfo
		{author} {\bibfnamefont {D.~M.}\ \bibnamefont {Burn}}, \bibinfo {author}
		{\bibfnamefont {G.}~\bibnamefont {van~der Laan}}, \bibinfo {author}
		{\bibfnamefont {T.}~\bibnamefont {Hesjedal}}, \bibinfo {author}
		{\bibfnamefont {Z.}~\bibnamefont {Yan}}, \bibinfo {author} {\bibfnamefont
			{J.}~\bibnamefont {Feng}}, \bibinfo {author} {\bibfnamefont {C.}~\bibnamefont
			{Wan}}, \bibinfo {author} {\bibfnamefont {J.}~\bibnamefont {Wei}}, \bibinfo
		{author} {\bibfnamefont {X.}~\bibnamefont {Wang}}, \bibinfo {author}
		{\bibfnamefont {X.}~\bibnamefont {Zhang}}, \bibinfo {author} {\bibfnamefont
			{H.}~\bibnamefont {Xu}}, \bibinfo {author} {\bibfnamefont {C.}~\bibnamefont
			{Guo}}, \bibinfo {author} {\bibfnamefont {H.}~\bibnamefont {Wei}}, \bibinfo
		{author} {\bibfnamefont {G.}~\bibnamefont {Finocchio}}, \bibinfo {author}
		{\bibfnamefont {X.}~\bibnamefont {Han}},\ and\ \bibinfo {author}
		{\bibfnamefont {G.}~\bibnamefont {Sch{\"u}tz}},\ }\bibfield  {title}
	{\bibinfo {title} {Anatomy of skyrmionic textures in magnetic multilayers},\
	}\href {https://doi.org/https://doi.org/10.1002/adma.201807683} {\bibfield
		{journal} {\bibinfo  {journal} {Adv. Mater.}\ }\textbf {\bibinfo {volume}
			{31}},\ \bibinfo {pages} {1807683} (\bibinfo {year} {2019})}\BibitemShut
	{NoStop}%
	\bibitem [{\citenamefont {Kim}\ \emph {et~al.}(2022)\citenamefont {Kim},
		\citenamefont {McCarter}, \citenamefont {Stoica}, \citenamefont {Das},
		\citenamefont {Klewe}, \citenamefont {Donoway}, \citenamefont {Burn},
		\citenamefont {Shafer}, \citenamefont {Rodolakis}, \citenamefont
		{Gonçalves}, \citenamefont {Gomez-Ortiz}, \citenamefont {Iniguez},
		\citenamefont {Garcia-Fernandez}, \citenamefont {Junquera}, \citenamefont
		{Susarla}, \citenamefont {Lovesey}, \citenamefont {van~der Laan},
		\citenamefont {Park}, \citenamefont {Martin}, \citenamefont {Freeland},
		\citenamefont {Ramesh},\ and\ \citenamefont {Lee}}]{Kim2022}%
	\BibitemOpen
	\bibfield  {author} {\bibinfo {author} {\bibfnamefont {K.~T.}\ \bibnamefont
			{Kim}}, \bibinfo {author} {\bibfnamefont {M.~R.}\ \bibnamefont {McCarter}},
		\bibinfo {author} {\bibfnamefont {V.~A.}\ \bibnamefont {Stoica}}, \bibinfo
		{author} {\bibfnamefont {S.}~\bibnamefont {Das}}, \bibinfo {author}
		{\bibfnamefont {C.}~\bibnamefont {Klewe}}, \bibinfo {author} {\bibfnamefont
			{E.~P.}\ \bibnamefont {Donoway}}, \bibinfo {author} {\bibfnamefont {D.~M.}\
			\bibnamefont {Burn}}, \bibinfo {author} {\bibfnamefont {P.}~\bibnamefont
			{Shafer}}, \bibinfo {author} {\bibfnamefont {F.}~\bibnamefont {Rodolakis}},
		\bibinfo {author} {\bibfnamefont {M.~A.~P.}\ \bibnamefont {Gonçalves}},
		\bibinfo {author} {\bibfnamefont {F.}~\bibnamefont {Gomez-Ortiz}}, \bibinfo
		{author} {\bibfnamefont {J.}~\bibnamefont {Iniguez}}, \bibinfo {author}
		{\bibfnamefont {P.}~\bibnamefont {Garcia-Fernandez}}, \bibinfo {author}
		{\bibfnamefont {J.}~\bibnamefont {Junquera}}, \bibinfo {author}
		{\bibfnamefont {S.}~\bibnamefont {Susarla}}, \bibinfo {author} {\bibfnamefont
			{S.~W.}\ \bibnamefont {Lovesey}}, \bibinfo {author} {\bibfnamefont
			{G.}~\bibnamefont {van~der Laan}}, \bibinfo {author} {\bibfnamefont {S.~Y.}\
			\bibnamefont {Park}}, \bibinfo {author} {\bibfnamefont {L.~W.}\ \bibnamefont
			{Martin}}, \bibinfo {author} {\bibfnamefont {J.~W.}\ \bibnamefont
			{Freeland}}, \bibinfo {author} {\bibfnamefont {R.}~\bibnamefont {Ramesh}},\
		and\ \bibinfo {author} {\bibfnamefont {D.~R.}\ \bibnamefont {Lee}},\
	}\bibfield  {title} {\bibinfo {title} {Chiral structures of electric
			polarization vectors quantified by x-ray resonant scattering},\ }\href
	{https://doi.org/10.1038/s41467-022-29359-5} {\bibfield  {journal} {\bibinfo
			{journal} {Nat. Commun.}\ }\textbf {\bibinfo {volume} {13}},\ \bibinfo
		{pages} {1769} (\bibinfo {year} {2022})}\BibitemShut {NoStop}%
	\bibitem [{\citenamefont {Arena}\ \emph {et~al.}(2009)\citenamefont {Arena},
		\citenamefont {Ding}, \citenamefont {Vescovo}, \citenamefont {Zohar},
		\citenamefont {Guan},\ and\ \citenamefont {Bailey}}]{Arena2009}%
	\BibitemOpen
	\bibfield  {author} {\bibinfo {author} {\bibfnamefont {D.~A.}\ \bibnamefont
			{Arena}}, \bibinfo {author} {\bibfnamefont {Y.}~\bibnamefont {Ding}},
		\bibinfo {author} {\bibfnamefont {E.}~\bibnamefont {Vescovo}}, \bibinfo
		{author} {\bibfnamefont {S.}~\bibnamefont {Zohar}}, \bibinfo {author}
		{\bibfnamefont {Y.}~\bibnamefont {Guan}},\ and\ \bibinfo {author}
		{\bibfnamefont {W.~E.}\ \bibnamefont {Bailey}},\ }\bibfield  {title}
	{\bibinfo {title} {{A compact apparatus for studies of element and
				phase-resolved ferromagnetic resonance}},\ }\href
	{https://doi.org/{10.1063/1.3190402}} {\bibfield  {journal} {\bibinfo
			{journal} {{Rev. Sci. Instrum.}}\ }\textbf {\bibinfo {volume} {{80}}},\
		\bibinfo {pages} {083903} (\bibinfo {year} {{2009}})}\BibitemShut {NoStop}%
	\bibitem [{\citenamefont {Marcham}\ \emph {et~al.}(2011)\citenamefont
		{Marcham}, \citenamefont {Keatley}, \citenamefont {Neudert}, \citenamefont
		{Hicken}, \citenamefont {Cavill}, \citenamefont {Shelford}, \citenamefont
		{van~der Laan}, \citenamefont {Telling}, \citenamefont {Childress},
		\citenamefont {Katine}, \citenamefont {Shafer},\ and\ \citenamefont
		{Arenholz}}]{Marcham2011}%
	\BibitemOpen
	\bibfield  {author} {\bibinfo {author} {\bibfnamefont {M.~K.}\ \bibnamefont
			{Marcham}}, \bibinfo {author} {\bibfnamefont {P.~S.}\ \bibnamefont
			{Keatley}}, \bibinfo {author} {\bibfnamefont {A.}~\bibnamefont {Neudert}},
		\bibinfo {author} {\bibfnamefont {R.~J.}\ \bibnamefont {Hicken}}, \bibinfo
		{author} {\bibfnamefont {S.~A.}\ \bibnamefont {Cavill}}, \bibinfo {author}
		{\bibfnamefont {L.~R.}\ \bibnamefont {Shelford}}, \bibinfo {author}
		{\bibfnamefont {G.}~\bibnamefont {van~der Laan}}, \bibinfo {author}
		{\bibfnamefont {N.~D.}\ \bibnamefont {Telling}}, \bibinfo {author}
		{\bibfnamefont {J.~R.}\ \bibnamefont {Childress}}, \bibinfo {author}
		{\bibfnamefont {J.~A.}\ \bibnamefont {Katine}}, \bibinfo {author}
		{\bibfnamefont {P.}~\bibnamefont {Shafer}},\ and\ \bibinfo {author}
		{\bibfnamefont {E.}~\bibnamefont {Arenholz}},\ }\bibfield  {title} {\bibinfo
		{title} {{Phase-resolved x-ray ferromagnetic resonance measurements in
				fluorescence yield}},\ }\href {https://doi.org/10.1063/1.3567143} {\bibfield
		{journal} {\bibinfo  {journal} {J. Appl. Phys.}\ }\textbf {\bibinfo {volume}
			{109}},\ \bibinfo {pages} {07D353} (\bibinfo {year} {2011})}\BibitemShut
	{NoStop}%
	\bibitem [{\citenamefont {Bailey}\ \emph {et~al.}(2013)\citenamefont {Bailey},
		\citenamefont {Cheng}, \citenamefont {Knut}, \citenamefont {Karis},
		\citenamefont {Auffret}, \citenamefont {Zohar}, \citenamefont {Keavney},
		\citenamefont {Warnicke}, \citenamefont {Lee},\ and\ \citenamefont
		{Arena}}]{bailey_XFMR}%
	\BibitemOpen
	\bibfield  {author} {\bibinfo {author} {\bibfnamefont {W.~E.}\ \bibnamefont
			{Bailey}}, \bibinfo {author} {\bibfnamefont {C.}~\bibnamefont {Cheng}},
		\bibinfo {author} {\bibfnamefont {R.}~\bibnamefont {Knut}}, \bibinfo {author}
		{\bibfnamefont {O.}~\bibnamefont {Karis}}, \bibinfo {author} {\bibfnamefont
			{S.}~\bibnamefont {Auffret}}, \bibinfo {author} {\bibfnamefont
			{S.}~\bibnamefont {Zohar}}, \bibinfo {author} {\bibfnamefont
			{D.}~\bibnamefont {Keavney}}, \bibinfo {author} {\bibfnamefont
			{P.}~\bibnamefont {Warnicke}}, \bibinfo {author} {\bibfnamefont {J.-S.}\
			\bibnamefont {Lee}},\ and\ \bibinfo {author} {\bibfnamefont {D.~A.}\
			\bibnamefont {Arena}},\ }\bibfield  {title} {\bibinfo {title} {Detection of
			microwave phase variation in nanometre-scale magnetic heterostructures},\
	}\href@noop {} {\bibfield  {journal} {\bibinfo  {journal} {Nat. Commun.}\
		}\textbf {\bibinfo {volume} {4}},\ \bibinfo {pages} {2025} (\bibinfo {year}
		{2013})}\BibitemShut {NoStop}%
	\bibitem [{\citenamefont {van~der Laan}(2017)}]{VanDerLaan_JESRP_2017}%
	\BibitemOpen
	\bibfield  {author} {\bibinfo {author} {\bibfnamefont {G.}~\bibnamefont
			{van~der Laan}},\ }\bibfield  {title} {\bibinfo {title} {{Time-resolved X-ray
				detected ferromagnetic resonance of spin currents}},\ }\href
	{https://doi.org/10.1016/j.elspec.2016.12.011} {\bibfield  {journal}
		{\bibinfo  {journal} {J. Electron. Spectrosc. Relat. Phenom.}\ }\textbf
		{\bibinfo {volume} {220}},\ \bibinfo {pages} {137} (\bibinfo {year}
		{2017})}\BibitemShut {NoStop}%
	\bibitem [{\citenamefont {Baker}\ \emph {et~al.}(2015)\citenamefont {Baker},
		\citenamefont {Figueroa}, \citenamefont {Collins-McIntyre}, \citenamefont
		{van~der Laan},\ and\ \citenamefont {Hesjedal}}]{Baker2015}%
	\BibitemOpen
	\bibfield  {author} {\bibinfo {author} {\bibfnamefont {A.~A.}\ \bibnamefont
			{Baker}}, \bibinfo {author} {\bibfnamefont {A.~I.}\ \bibnamefont {Figueroa}},
		\bibinfo {author} {\bibfnamefont {L.~J.}\ \bibnamefont {Collins-McIntyre}},
		\bibinfo {author} {\bibfnamefont {G.}~\bibnamefont {van~der Laan}},\ and\
		\bibinfo {author} {\bibfnamefont {T.}~\bibnamefont {Hesjedal}},\ }\bibfield
	{title} {\bibinfo {title} {{Spin pumping in Ferromagnet-Topological
				Insulator-Ferromagnet Heterostructures}},\ }\href
	{https://doi.org/10.1038/srep07907} {\bibfield  {journal} {\bibinfo
			{journal} {Sci. Rep.}\ }\textbf {\bibinfo {volume} {5}},\ \bibinfo {pages}
		{7907} (\bibinfo {year} {2015})}\BibitemShut {NoStop}%
	\bibitem [{\citenamefont {Baker}\ \emph {et~al.}(2016)\citenamefont {Baker},
		\citenamefont {Figueroa}, \citenamefont {Pingstone}, \citenamefont {Lazarov},
		\citenamefont {van~der Laan},\ and\ \citenamefont {Hesjedal}}]{Baker2016}%
	\BibitemOpen
	\bibfield  {author} {\bibinfo {author} {\bibfnamefont {A.~A.}\ \bibnamefont
			{Baker}}, \bibinfo {author} {\bibfnamefont {A.~I.}\ \bibnamefont {Figueroa}},
		\bibinfo {author} {\bibfnamefont {D.}~\bibnamefont {Pingstone}}, \bibinfo
		{author} {\bibfnamefont {V.~K.}\ \bibnamefont {Lazarov}}, \bibinfo {author}
		{\bibfnamefont {G.}~\bibnamefont {van~der Laan}},\ and\ \bibinfo {author}
		{\bibfnamefont {T.}~\bibnamefont {Hesjedal}},\ }\bibfield  {title} {\bibinfo
		{title} {{Spin pumping in magnetic trilayer structures with an MgO
				barrier}},\ }\href {https://doi.org/10.1038/srep35582} {\bibfield  {journal}
		{\bibinfo  {journal} {Sci. Rep.}\ }\textbf {\bibinfo {volume} {6}},\ \bibinfo
		{pages} {35582} (\bibinfo {year} {2016})}\BibitemShut {NoStop}%
	\bibitem [{\citenamefont {Burn}\ \emph
		{et~al.}(2020{\natexlab{a}})\citenamefont {Burn}, \citenamefont {Zhang},
		\citenamefont {Zhai}, \citenamefont {Chai}, \citenamefont {Sun},
		\citenamefont {van~der Laan},\ and\ \citenamefont
		{Hesjedal}}]{Burn2020-DFMR}%
	\BibitemOpen
	\bibfield  {author} {\bibinfo {author} {\bibfnamefont {D.~M.}\ \bibnamefont
			{Burn}}, \bibinfo {author} {\bibfnamefont {S.}~\bibnamefont {Zhang}},
		\bibinfo {author} {\bibfnamefont {K.}~\bibnamefont {Zhai}}, \bibinfo {author}
		{\bibfnamefont {Y.}~\bibnamefont {Chai}}, \bibinfo {author} {\bibfnamefont
			{Y.}~\bibnamefont {Sun}}, \bibinfo {author} {\bibfnamefont {G.}~\bibnamefont
			{van~der Laan}},\ and\ \bibinfo {author} {\bibfnamefont {T.}~\bibnamefont
			{Hesjedal}},\ }\bibfield  {title} {\bibinfo {title} {Mode-resolved detection
			of magnetization dynamics using x-ray diffractive ferromagnetic resonance},\
	}\href {https://doi.org/10.1021/acs.nanolett.9b03989} {\bibfield  {journal}
		{\bibinfo  {journal} {Nano Lett.}\ }\textbf {\bibinfo {volume} {20}},\
		\bibinfo {pages} {345} (\bibinfo {year} {2020}{\natexlab{a}})}\BibitemShut
	{NoStop}%
	\bibitem [{\citenamefont {Burn}\ \emph
		{et~al.}(2020{\natexlab{b}})\citenamefont {Burn}, \citenamefont {Zhang},
		\citenamefont {Yu}, \citenamefont {Guang}, \citenamefont {Chen},
		\citenamefont {Qiu}, \citenamefont {van~der Laan},\ and\ \citenamefont
		{Hesjedal}}]{PhysRevLett.125.137201}%
	\BibitemOpen
	\bibfield  {author} {\bibinfo {author} {\bibfnamefont {D.~M.}\ \bibnamefont
			{Burn}}, \bibinfo {author} {\bibfnamefont {S.~L.}\ \bibnamefont {Zhang}},
		\bibinfo {author} {\bibfnamefont {G.~Q.}\ \bibnamefont {Yu}}, \bibinfo
		{author} {\bibfnamefont {Y.}~\bibnamefont {Guang}}, \bibinfo {author}
		{\bibfnamefont {H.~J.}\ \bibnamefont {Chen}}, \bibinfo {author}
		{\bibfnamefont {X.~P.}\ \bibnamefont {Qiu}}, \bibinfo {author} {\bibfnamefont
			{G.}~\bibnamefont {van~der Laan}},\ and\ \bibinfo {author} {\bibfnamefont
			{T.}~\bibnamefont {Hesjedal}},\ }\bibfield  {title} {\bibinfo {title}
		{Depth-resolved magnetization dynamics revealed by x-ray reflectometry
			ferromagnetic resonance},\ }\href
	{https://doi.org/10.1103/PhysRevLett.125.137201} {\bibfield  {journal}
		{\bibinfo  {journal} {Phys. Rev. Lett.}\ }\textbf {\bibinfo {volume} {125}},\
		\bibinfo {pages} {137201} (\bibinfo {year} {2020}{\natexlab{b}})}\BibitemShut
	{NoStop}%
	\bibitem [{\citenamefont {P\"ollath}\ \emph {et~al.}(2019)\citenamefont
		{P\"ollath}, \citenamefont {Aqeel}, \citenamefont {Bauer}, \citenamefont
		{Luo}, \citenamefont {Ryll}, \citenamefont {Radu}, \citenamefont
		{Pfleiderer}, \citenamefont {Woltersdorf},\ and\ \citenamefont
		{Back}}]{Back-DFMR}%
	\BibitemOpen
	\bibfield  {author} {\bibinfo {author} {\bibfnamefont {S.}~\bibnamefont
			{P\"ollath}}, \bibinfo {author} {\bibfnamefont {A.}~\bibnamefont {Aqeel}},
		\bibinfo {author} {\bibfnamefont {A.}~\bibnamefont {Bauer}}, \bibinfo
		{author} {\bibfnamefont {C.}~\bibnamefont {Luo}}, \bibinfo {author}
		{\bibfnamefont {H.}~\bibnamefont {Ryll}}, \bibinfo {author} {\bibfnamefont
			{F.}~\bibnamefont {Radu}}, \bibinfo {author} {\bibfnamefont {C.}~\bibnamefont
			{Pfleiderer}}, \bibinfo {author} {\bibfnamefont {G.}~\bibnamefont
			{Woltersdorf}},\ and\ \bibinfo {author} {\bibfnamefont {C.~H.}\ \bibnamefont
			{Back}},\ }\bibfield  {title} {\bibinfo {title} {Ferromagnetic resonance with
			magnetic phase selectivity by means of resonant elastic x-ray scattering on a
			chiral magnet},\ }\href {https://doi.org/10.1103/PhysRevLett.123.167201}
	{\bibfield  {journal} {\bibinfo  {journal} {Phys. Rev. Lett.}\ }\textbf
		{\bibinfo {volume} {123}},\ \bibinfo {pages} {167201} (\bibinfo {year}
		{2019})}\BibitemShut {NoStop}%
	\bibitem [{\citenamefont {Maksymov}\ and\ \citenamefont
		{Kostylev}(2015)}]{BroadbandFMR}%
	\BibitemOpen
	\bibfield  {author} {\bibinfo {author} {\bibfnamefont {I.~S.}\ \bibnamefont
			{Maksymov}}\ and\ \bibinfo {author} {\bibfnamefont {M.}~\bibnamefont
			{Kostylev}},\ }\bibfield  {title} {\bibinfo {title} {Broadband stripline
			ferromagnetic resonance spectroscopy of ferromagnetic films, multilayers and
			nanostructures},\ }\href
	{https://doi.org/https://doi.org/10.1016/j.physe.2014.12.027} {\bibfield
		{journal} {\bibinfo  {journal} {Physica E}\ }\textbf {\bibinfo {volume}
			{69}},\ \bibinfo {pages} {253} (\bibinfo {year} {2015})}\BibitemShut
	{NoStop}%
	\bibitem [{\citenamefont {Templeton}\ and\ \citenamefont
		{Templeton}(1994)}]{Templeton1994}%
	\BibitemOpen
	\bibfield  {author} {\bibinfo {author} {\bibfnamefont {D.~H.}\ \bibnamefont
			{Templeton}}\ and\ \bibinfo {author} {\bibfnamefont {L.~K.}\ \bibnamefont
			{Templeton}},\ }\bibfield  {title} {\bibinfo {title} {Tetrahedral anisotropy
			of x-ray anomalous scattering},\ }\href
	{https://doi.org/10.1103/PhysRevB.49.14850} {\bibfield  {journal} {\bibinfo
			{journal} {Phys. Rev. B}\ }\textbf {\bibinfo {volume} {49}},\ \bibinfo
		{pages} {14850} (\bibinfo {year} {1994})}\BibitemShut {NoStop}%
	\bibitem [{\citenamefont {Dmitrienko}\ and\ \citenamefont
		{Chizhikov}(2012)}]{Dmitrienko2012}%
	\BibitemOpen
	\bibfield  {author} {\bibinfo {author} {\bibfnamefont {V.~E.}\ \bibnamefont
			{Dmitrienko}}\ and\ \bibinfo {author} {\bibfnamefont {V.~A.}\ \bibnamefont
			{Chizhikov}},\ }\bibfield  {title} {\bibinfo {title} {Weak antiferromagnetic
			ordering induced by dzyaloshinskii-moriya interaction and pure magnetic
			reflections in mnsi-type crystals},\ }\href
	{https://doi.org/10.1103/PhysRevLett.108.187203} {\bibfield  {journal}
		{\bibinfo  {journal} {Phys. Rev. Lett.}\ }\textbf {\bibinfo {volume} {108}},\
		\bibinfo {pages} {187203} (\bibinfo {year} {2012})}\BibitemShut {NoStop}%
	\bibitem [{\citenamefont {{van der Laan}}(2008)}]{VANDERLAAN2008570}%
	\BibitemOpen
	\bibfield  {author} {\bibinfo {author} {\bibfnamefont {G.}~\bibnamefont {{van
					der Laan}}},\ }\bibfield  {title} {\bibinfo {title} {{Soft X-ray resonant
				magnetic scattering of magnetic nanostructures}},\ }\href
	{https://doi.org/https://doi.org/10.1016/j.crhy.2007.06.004} {\bibfield
		{journal} {\bibinfo  {journal} {C. R. Physique}\ }\textbf {\bibinfo {volume}
			{9}},\ \bibinfo {pages} {570} (\bibinfo {year} {2008})}\BibitemShut {NoStop}%
	\bibitem [{\citenamefont {Gurevich}\ and\ \citenamefont
		{Melkov}(1996)}]{Gurevich}%
	\BibitemOpen
	\bibfield  {author} {\bibinfo {author} {\bibfnamefont {A.~G.}\ \bibnamefont
			{Gurevich}}\ and\ \bibinfo {author} {\bibfnamefont {G.~A.}\ \bibnamefont
			{Melkov}},\ }\href@noop {} {\emph {\bibinfo {title} {Magnetization
				Oscillations and Waves}}}\ (\bibinfo  {publisher} {CRC Press},\ \bibinfo
	{address} {Boca Raton},\ \bibinfo {year} {1996})\BibitemShut {NoStop}%
	\bibitem [{\citenamefont {Seki}\ \emph
		{et~al.}(2012{\natexlab{a}})\citenamefont {Seki}, \citenamefont {Kim},
		\citenamefont {Inosov}, \citenamefont {Georgii}, \citenamefont {Keimer},
		\citenamefont {Ishiwata},\ and\ \citenamefont {Tokura}}]{Seki_PRB_2012}%
	\BibitemOpen
	\bibfield  {author} {\bibinfo {author} {\bibfnamefont {S.}~\bibnamefont
			{Seki}}, \bibinfo {author} {\bibfnamefont {J.~H.}\ \bibnamefont {Kim}},
		\bibinfo {author} {\bibfnamefont {D.~S.}\ \bibnamefont {Inosov}}, \bibinfo
		{author} {\bibfnamefont {R.}~\bibnamefont {Georgii}}, \bibinfo {author}
		{\bibfnamefont {B.}~\bibnamefont {Keimer}}, \bibinfo {author} {\bibfnamefont
			{S.}~\bibnamefont {Ishiwata}},\ and\ \bibinfo {author} {\bibfnamefont
			{Y.}~\bibnamefont {Tokura}},\ }\bibfield  {title} {\bibinfo {title}
		{{Formation and rotation of skyrmion crystal in the chiral-lattice insulator
				Cu$_2$OSeO$_3$}},\ }\href {https://doi.org/10.1103/PhysRevB.85.220406}
	{\bibfield  {journal} {\bibinfo  {journal} {Phys. Rev. B}\ }\textbf {\bibinfo
			{volume} {85}},\ \bibinfo {pages} {220406} (\bibinfo {year}
		{2012}{\natexlab{a}})}\BibitemShut {NoStop}%
	\bibitem [{\citenamefont {Seki}\ \emph
		{et~al.}(2012{\natexlab{b}})\citenamefont {Seki}, \citenamefont {Yu},
		\citenamefont {Ishiwata},\ and\ \citenamefont {Tokura}}]{Seki_Science_2012}%
	\BibitemOpen
	\bibfield  {author} {\bibinfo {author} {\bibfnamefont {S.}~\bibnamefont
			{Seki}}, \bibinfo {author} {\bibfnamefont {X.~Z.}\ \bibnamefont {Yu}},
		\bibinfo {author} {\bibfnamefont {S.}~\bibnamefont {Ishiwata}},\ and\
		\bibinfo {author} {\bibfnamefont {Y.}~\bibnamefont {Tokura}},\ }\bibfield
	{title} {\bibinfo {title} {{Observation of Skyrmions in a Multiferroic
				Material}},\ }\href {https://doi.org/10.1126/science.1214143} {\bibfield
		{journal} {\bibinfo  {journal} {Science}\ }\textbf {\bibinfo {volume}
			{336}},\ \bibinfo {pages} {198} (\bibinfo {year}
		{2012}{\natexlab{b}})}\BibitemShut {NoStop}%
	\bibitem [{\citenamefont {Storey}\ \emph {et~al.}(1977)\citenamefont {Storey},
		\citenamefont {Tooke}, \citenamefont {Cracknell},\ and\ \citenamefont
		{Przystawa}}]{Storey_1977}%
	\BibitemOpen
	\bibfield  {author} {\bibinfo {author} {\bibfnamefont {B.~E.}\ \bibnamefont
			{Storey}}, \bibinfo {author} {\bibfnamefont {A.~O.}\ \bibnamefont {Tooke}},
		\bibinfo {author} {\bibfnamefont {A.~P.}\ \bibnamefont {Cracknell}},\ and\
		\bibinfo {author} {\bibfnamefont {J.~A.}\ \bibnamefont {Przystawa}},\
	}\bibfield  {title} {\bibinfo {title} {The determination of the frequencies
			of magnetostatic modes in rectangular thin films of ferrimagnetic yttrium
			iron garnet},\ }\href {https://doi.org/10.1088/0022-3719/10/6/017} {\bibfield
		{journal} {\bibinfo  {journal} {J. Phys. C: Solid State Phys.}\ }\textbf
		{\bibinfo {volume} {10}},\ \bibinfo {pages} {875} (\bibinfo {year}
		{1977})}\BibitemShut {NoStop}%
	\bibitem [{sup()}]{suppl}%
	\BibitemOpen
	\href@noop {} {\bibinfo {title} {{See Supplemental Material at [URL will be
				inserted by publisher] for an animated illustration of the LH conical screw
				in a positive applied field (Static{\_}+Q{\_}+H.mp4) and of the dynamic
				conical spin structures (for $+Q$) in applied fields along the helical
				propagation axis and opposite to it (Dynamics{\_}+Q{\_}+H.mp4 and
				Dynamics{\_}+Q{\_}-H.mp4, respectively)}}}\BibitemShut {NoStop}%
\end{thebibliography}

%

\end{document}